\documentclass[usenatbib]{mn2e}

\usepackage{times}

\usepackage{graphicx}
\graphicspath{{./Pics/}}

\newcommand{\un}[1]{{\,{\rm #1}}}

\title[Meridional flows, butterfly diagrams, and polar caps]{The 
impact of meridional circulation on stellar butterfly diagrams and
polar caps}

\author[V.\ Holzwarth, D.\ H.\ Mackay, M.\ Jardine]{V. Holzwarth$^{1}$
\thanks{E-mail: vrh1@st-andrews.ac.uk}, D.~H.~Mackay$^{2}$, M.
Jardine$^{1}$ \\
$^1$School of Physics and Astronomy, University of St Andrews, 
North Haugh, St Andrews, Fife KY16 9SS, Scotland \\
$^2$School of Mathematics and Statistics, University of St Andrews, 
North Haugh, St Andrews, Fife KY16 9SS, Scotland
}

\begin{document}

\date{Received; accepted 2005}

\pagerange{\pageref{firstpage}--\pageref{lastpage}} \pubyear{2002}

\maketitle

\label{firstpage}

\begin{abstract} 
Observations of rapidly rotating solar-like stars show a significant 
mixture of opposite-polarity magnetic fields within their polar
regions.
To explain these observations, models describing the surface transport
of magnetic flux demand the presence of fast meridional flows.
Here, we link sub-surface and surface magnetic flux transport
simulations to investigate (i) the impact of meridional circulations
with peak velocities of $\le 125\un{m\cdot s^{-1}}$ on the latitudinal
eruption pattern of magnetic flux tubes and (ii) the influence of the
resulting butterfly diagrams on polar magnetic field properties.
Prior to their eruption, magnetic flux tubes with low field strengths 
and initial cross sections below $\sim 300\un{km}$ experience an 
enhanced poleward deflection through meridional flows (assumed to be 
poleward at the top of the convection zone and equatorward at the
bottom).
In particular flux tubes which originate between low and intermediate
latitudes within the convective overshoot region are strongly affected.
This latitude-dependent poleward deflection of erupting magnetic flux
renders the wings of stellar butterfly diagrams distinctively convex.
The subsequent evolution of the surface magnetic field shows that the 
increased number of newly emerging bipoles at higher latitudes promotes
the intermingling of opposite polarities of polar magnetic fields.
The associated magnetic flux densities are about $20\%$ higher than in
the case disregarding the pre-eruptive deflection, which eases the
necessity for fast meridional flows predicted by previous
investigations.
In order to reproduce the observed polar field properties, the rate of 
the meridional circulation has to be on the order of $100\un{m\cdot 
s^{-1}}$, and the latitudinal range from which magnetic flux tubes 
originate at the base of the convective zone ($\la 50\degr$) must be 
larger than in the solar case ($\la 35\degr$).
\end{abstract}

\begin{keywords}
 stars: magnetic fields --
 stars: activity -- 
 stars: interior --
 stars: rotation --  
 stars: spots --
 stars: imaging
\end{keywords}

\section{Introduction}
\label{intro}
On the Sun, dark spots are exclusively found within an equatorial belt 
between about $\pm 40\degr$ latitude.
In contrast, stellar surface brightness maps, secured with the 
technique of Doppler imaging \citep[ and references
therein]{2001astr.conf..183C}, show that other cool stars frequently
have large high-latitude and polar spots, often in conjunction with
low-latitude features as well \citep[ and references
therein]{2002AN....323..309S}.
Due to specific requirements of the observing technique the targets so 
far are stars rotating more rapidly than the Sun.
Although the time base of surface maps acquired for individual stars is
yet not sufficiently long to conclusively discern long-term activity
properties, their behaviour seems to digress from the solar 11-year 
spot/22-year magnetic cycles and polar field properties.
In fact, many of the younger and more active stars show no apparent 
cycle insofar as chromospheric indicators such as Ca\,II H \& K can be 
used as a proxy for magnetic activity \citep{1995ApJ...438..269B,
Donahue1996}.
There is, as yet, also no example of a star undergoing a `Maunder
minimum'.
Different activity and cycle signatures may however be present or more
pronounced, such as apparent preferred longitudes in the spot
distribution or the `flip-flop' phenomenon \citep{1994A&A...282L...9J, 
2004SoPh..224..123B, 2004MNRAS.352L..17M, 2005A&A...440.1161K}.

An important characteristic difference to the solar-like magnetic field
distribution is the significant mixture of magnetic flux of opposite 
polarity within polar regions, as observed with the Zeeman-Doppler
Imaging technique \citep{semel89, donati97}.
In particular the strong flux intermingling in the polar regions is in
contrast to the Sun \citep{2003MNRAS.345.1145D}, where the
high-latitude magnetic field is essentially unipolar throughout the
majority of the activity cycle.
On rapidly rotating stars, the associated polar magnetic flux densities
are sufficiently high and persistent to cause dark polar caps lasting
over a large number of stellar rotation periods
\citep[e.g.][]{jeffers2005abdor}.

The magnetic activity signatures of cool stars are ascribed to the
emergence of magnetic flux generated by sub-surface dynamo mechanisms.
Scenarios for dynamo operation are based on the magneto-hydrodynamic
interaction between convective motions and (differential) rotation.
Yet there is no complete theory, which unifies the amplification, 
storage, transport, eruption, and (possibly cyclic) re-generation of 
magnetic flux consistently; for an extensive review on stellar dynamo
theory see \citet{2003A&ARv..11..287O}.
Dynamo mechanisms inside the convective envelope leave characteristic
imprints on the observable activity signatures in the stellar
atmosphere, which provide constraints for the underlying processes.

In the following, we focus on the transport of magnetic flux both below
and on the stellar surface.
Magnetic flux tube models \citep[e.g.][]{1986A&A...166..291M,
1987ApJ...316..788C, 1993A&A...272..621D, 1994ApJ...436..907F,
1994A&A...281L..69S, 1996A&A...314..503S} describe the evolution of 
magnetic flux, concentrated in strands of magnetic field lines, inside 
the convection zone until their eruption on the stellar surface.
Although they exclude aspects concerning the (cyclic) generation of
magnetic fields, they successfully reproduce characteristic properties 
of emerging bi-polar spot groups like, in the case of the Sun, their 
latitude of emergence, their relative velocities and topologies, the
asymmetries between the preceding- and following spot group, and Joy's
law \citep[e.g.][]{1993A&A...272..621D, 1995ApJ...441..886C}.
Further applications of the flux eruption model comprise cool stars 
with different rotation rates, stellar masses, and evolutionary stages 
as well as components of close binary systems 
\citep{2000A&A...355.1087G, 2001A&A...377..251H, 2003A&A...405..303H,
2004AN....325..408H}.
Frequently observed starspots at higher latitudes of rapid rotators,
for example, can be explained by the poleward deflection of rising flux
tubes prior to their eruption on the stellar surface
\citep[e.g.][]{1992A&A...264L..13S}.
A persistent magnetic flux eruption at high latitudes implies however 
the availability (and possibly generation) of large amounts of magnetic
flux over a similar latitudinal range inside the stellar convection
zone, which would be in dissent with the solar case, where the
production of magnetic flux is anticipated to be most efficient at low
latitudes \citep[e.g.][]{1999SoPh..184...61C}.

Surface flux transport models \citep[e.g.][]{DeVore1984, Wang1989a, 
vanBal1998, Schrijver2001b, Mackay2002b, Mackay2004, 
2004A&A...426.1075B} follow the evolution of the radial magnetic field
component on the stellar surface under the combined effects of magnetic
flux emergence, differential rotation, meridional flow, and
supergranular diffusion.
Using empirical properties of bi-polar regions (i.e.\ latitudinal
migration of emergence rates and tilt angles) they successfully
reproduce major features of the solar cycle like the reversal of the
polar field.
Considering moderately rotating stars, \citet{Schrijver2001b} showed 
that an enhancement of the rate of flux emergence produces dark polar
caps of a single magnetic polarity, surrounded by a flux ring of
opposite polarity.
More recently, \citet{Mackay2004} showed that additional enhancements 
of both the latitudinal range of flux emergence and the meridional flow
velocity are required to successfully reproduce a significant mixture
of magnetic polarities at high latitudes on rapid rotators.

The surface flux transport models may accurately describe the surface
evolution of the magnetic field, but they do not address the question
of whether the assumed properties of erupting magnetic flux are
consistent with the requirement of high meridional flow velocities.
In particular, how the latitudinal distributions of flux eruption are
affected by enhanced rates of meridional flows.
We therefore link our studies on the pre-eruptive and the post-eruptive
evolution of magnetic flux to consistently quantify the impact of
meridional circulations on the butterfly diagram and the polar magnetic
field properties of rapidly rotating stars.
The main aim of this investigation is to validate the assumptions of 
\citet{Mackay2004} about extended latitudinal ranges of magnetic flux
emergence.
Yet the relation between observable activity properties and sub-surface
transport mechanisms also makes it possible to infer empirical 
constraints for specific dynamo properties.
In Sect.\ \ref{erupt}, we investigate the rise of magnetic flux tubes
through the convection zone prior to their emergence on the stellar
surface to determine how the latitudinal eruption pattern depends on
the strength of the circulation.
The results are used in Sect.\ \ref{butter} to determine the impact of
meridional circulations on stellar butterfly diagrams.
In Sect.\ \ref{surft}, we use the specified flux emergence latitudes 
within surface flux transport simulations to determine the requirements
for the reproduction of observed stellar magnetic field properties.
In Sect.\ \ref{disc}, we discuss our results and their implications for
possible dynamo scenarios.
Our conclusions are summarised in Sect.\ \ref{conc}.

\section{Sub-surface evolution and eruption of magnetic flux tubes}
\label{erupt}

\subsection{Basic scenario}
Magnetic activity signatures in the atmosphere of cool stars are
ascribed to the eruption of magnetic flux tubes, which are generated by
sub-surface dynamo processes inside the convective envelope
\citep[e.g.][ and references therein]{2003A&ARv..11..287O,
2005AN....326..194S}.
The amplification of the magnetic field is expected to take place at 
the tachocline, a region of strong shear flows at the interface to the 
radiative core of the star.
Helioseismological observations indicate that the tachocline is
slightly prolate \citep[e.g.][]{1999ApJ...527..445C, 
2001MNRAS.324..498B}.
At the equator, the bulk of the tachocline is right beneath the 
convection zone, whereas at higher latitudes a substantial part of it
is located inside the convection zone.
In the dynamically unstable stratification of the convection zone,
magnetic flux is subject to magnetic buoyancy, which leads to its rapid 
loss through eruption \citep{1975ApJ...198..205P, 1983A&A...122..241M}.
In the case of the Sun, for example, a magnetic flux tube can traverse 
the convection zone within several weeks.
The generation of high field strengths requires magnetic fields to 
persist in the amplifying region over time scales comparable with those
of supposed dynamo processes.
This requirement leads to the conjecture of magnetic flux being stored
inside the subadiabatic overshoot region beneath the convection zone
\citep{1982A&A...106...58S, 1982A&A...113...99V}.
Inside this stably stratified region, magnetic flux tubes are perturbed 
through overshooting gas plumes penetrating from the convective 
envelope above.
If its magnetic field strength is sufficiently large, a displaced flux 
tube is liable to a buoyancy-driven instability 
\citep[cf.][]{1966ApJ...145..811P, 1993A&A...272..621D,
1994A&A...281L..69S}.
Once an unstable, growing flux loop is properly located inside the
superadiabatic convection zone, it rapidly rises to the stellar
surface.
Coriolis forces, induced by the internal plasma flow along the flux
tube, cause an asymmetric evolution of the proceeding- and following
legs of the tube (relative to the direction of stellar rotation), which
twists the upper part of the rising loop
\citep[e.g.][]{1995ApJ...441..886C}.
Upon eruption on the stellar photosphere, this twists entails a tilt
between the polarity centres of the emerging bipole and the parallels
on the stellar surface.
Depending on their size and average field strength, the erupted flux 
tubes produce a spectrum of magnetic activity signatures, from small 
magnetic knots and pores to bi-polar spot groups and active regions
\citep[e.g.][]{2000sostact.book..S}.
Newly emerged surface magnetic flux dynamically disconnects from the
sub-surface parts of the parent flux tube \citep{1999SoPh..188..331S,
2005A&A...441..337S}; in the case of the Sun this process is expected
to take place after a few days in a depth less than $10\un{Mm}$ below
the surface.
The associated upflow of entropy-rich plasma inside the decapitated 
flux tube, from its anchor rooted in the overshoot region into the 
upper convection zone, weakens the magnetic field of the tube stumps 
and promotes their disintegration through turbulent diffusion driven by
magneto-convective motions.
The dissolving magnetic field may be transported downwards to the 
stable overshoot region through meridional circulation and convective 
pumping \citep[e.g.][]{2001ApJ...549.1183T}, providing the seed field 
for further amplification or a new activity cycle.

There is currently no detailed model for the removal of `old' flux from
the lower convection zone prior to the production of magnetic flux of
the opposite polarity in the successive cycle.
Neither it is clear how, if at all, the transformation of toroidal to
poloidal magnetic fields through the twisting of erupting flux loops is
related to the re-generation of magnetic flux at the bottom of the 
convection zone.
The lack of models prevents a prediction of the amount and location of 
magnetic flux as a function of the meridional flow pattern and 
-velocities.
Focusing on the transport of magnetic flux only, we hence presume the
existence of appropriate magnetic flux tubes at the bottom of the
convection zone of rapidly rotating stars, without further assumptions
about their amplification and cyclic re-generation.

\subsection{Model setup}

\subsubsection{Thin magnetic flux tubes}
The first part of the investigation is carried out in the framework of
the thin flux tube approximation \citep{1981A&A...102..129S}.
For this approximation to be applicable, the radius of the flux tube
must be smaller than all other relevant length scales like, for
example, the scale height of the pressure or the superadiabaticity, the
local radius of curvature of the magnetic field, or the wavelength of
perturbations propagating along the tube.
Assuming an ideal plasma with infinite conductivity, the magnetic flux
over a tube's circular cross section is conserved.

Due to the relatively short signal timescale across the tube's 
diameter, the magnetic flux tube is in instantaneous pressure
equilibrium with its environment, with the sum of the gas and magnetic 
pressure inside the tube balancing the gas pressure of the field-free 
external plasma, 
\begin{equation}
p_{e}= p + \frac{B^2}{8\pi}
\ .
\label{pressequi}
\end{equation}
The total pressure changes smoothly across the tube's surface
\citep[e.g.][]{1989sun.book.....S}.
For magnetic fields inside the convection zone the magnetic pressure is
typically smaller than the gas pressure (i.e.\ $B^2/(8\pi)\ll p$).
Since the flux tube is impenetrable for the external plasma,
perpendicular motions relative to the tube's axis cause the external 
plasma to flow around the tube.
The reaction of this distortion on the flux tube dynamics is quantified
in terms of a hydrodynamic drag,
\begin{equation}
\vec{f}_D
=
\rho_e 
\frac{C_D}{\pi a} \left| v_\perp \right| \vec{v}_\perp
\ ,
\label{fdrag}
\end{equation}
with $\rho_e$ being the density of the external medium.
The drag force tries to reduce the perpendicular velocity difference,
$\vec{v}_\perp= v_\perp \vec{e}_\perp= \left( \vec{v}_{e} - \vec{v} 
\right)_\perp$, between the external and internal motions, $\vec{v}_e$
and $\vec{v}$, respectively.
The value of the dimensionless drag coefficient, $C_D$, depends on the 
radius, $a$, of the tube considered in relation to the spectrum of
length scales of the convective motions in its vicinity.
The value of the effective turbulent viscosity depends on the specific
model description of the convective energy transport.
Since it is typically much larger than the molecular viscosity, the 
drag coefficient is anticipated to be of order one.

In addition to the interaction with its environment, realised through
the lateral pressure equilibrium (i.e.\ magnetic buoyancy) and the
hydrodynamic drag, the evolution of a magnetic flux tube also depends
on the magnetic curvature force and the Coriolis force.
In rapidly rotating stars, the interplay between the last two forces
typically causes a deflection of rising flux loops to higher latitudes
\citep[e.g.][]{1987ApJ...316..788C, 1992A&A...264L..13S,
1995ApJ...441..886C}.

\subsubsection{Stellar stratification and meridional circulation}
We consider the evolution of magnetic flux tubes in a $1\un{M_{\sun}}$ 
star, described by a spherically symmetric model of the current Sun
(i.e.\ $R_\star= R_{\sun}= 6.96\cdot10^{10}\un{cm}$).
The outer convection zone extends down to about $0.72\un{R_{\sun}}$ and 
comprises at its lower boundary a superadiabatically stratified
overshoot region of about $10^4\un{km}$ depth.
The stellar rotation period is taken to be $6\un{d}$ ($\Omega=
1.212\cdot10^{-5}\un{s^{-1}}$); for the sake of simplicity differential 
rotation is neglected.

The meridional flow pattern, $\vec{v}_p$, is superposed on the stellar 
structure, anticipating that its influence on the underlying
hydrostatic stratification is negligibly small.
We follow the approach of \citet{1988ApJ...333..965V} and describe the 
meridional circulation analytically through the poloidal velocity
components
\begin{eqnarray}
v_{p,r} 
& = &
u_0
\left( 1 + \xi \right)^2
F
\sin^m \theta
\left[ \left( m+2 \right) \cos^2 \theta - \sin^2 \theta \right]
\label{vpr}
\\
v_{p,\theta} 
& = &
-
u_0
\frac{G}{\sin \theta} 
\left( 1 + \xi \right)^3
\left( 1 - c_1 \xi^n + c_2 \xi^{n+k} \right)
\ ,
\label{vptheta}
\end{eqnarray}
where $F (\xi) $ and $G (\theta)$ are univariate functions of the 
radius, $r$, and the co-latitude, $\theta$, respectively, with $\xi= 
R_\star / r - 1$; a more detailed description of this flow model is 
given in Appdx.\ \ref{meriflow}.
The circulation is parametrised through the location of its lower
boundary, $r_b$, and the dimensionless quantities $n$, $m$, and $k$,
which also define the coefficients $c_1$ and $c_2$ in Eqs.\ 
(\ref{c1coeff}) and (\ref{c2coeff}), respectively.
We locate the maximum of the latitudinal flow velocity at co-latitude
$\theta_M= 53\degr$ to obtain a solar-like surface flow.
Since the radial profiles of stellar meridional circulations are as yet
virtually unknown, we adopt for the radial parameters $r_b$ and $k$ the
values used by \citet{1988ApJ...333..965V}.
Thus, the assumed meridional flow pattern is quantified through the set
of parameters $r_b= 0.7\un{R_\star}$, $n= 1.5$, $k= 0.5$, and $m=
0.76$.
It is composed of a single-cell circulation (per hemisphere) with a 
poleward flow at the stellar surface and an equatorward flow at the 
bottom of the convection zone (Fig.\ \ref{lattraj.fig}).
The amplitude of the circulation, $u_0\approx - 2.47\, v_M$ [cf.\ Eq.\
(\ref{defu0})], depends on the choice of the peak flow velocity, $v_M$,
on the stellar surface.

\subsubsection{Flux tube equilibria}
\label{fteq}
The magnetic flux tubes start their evolution from a consistently
defined initial configuration.
For this we assume a mechanical equilibrium, which implies that their
orientation is parallel to the equatorial plane and that their radius 
of curvature, $R_0= r_0 \cos \lambda_0$, is constant, where $r_0$ and
$\lambda_0$ are the equilibrium radius and latitude, respectively.
In the absence of meridional flows, flux rings in mechanical equilibrium
are non-buoyant and the magnetic curvature force (pointing toward the 
axis of stellar rotation) is balanced by inertia and Coriolis forces 
(pointing away from the rotation axis).
The Coriolis force is caused by a plasma flow inside the flux tube
prograde to the stellar rotation.
The properties of the mechanical equilibrium are described 
by \citet{1982A&A...106...58S, 1993GAFD...72..209, 1995GAFD...81..233};
for different approaches see \citet{1989SoPh..123..217C,
1992A&A...264..686M, 1994ApJ...436..907F, 1998ApJ...502..481C,
2003A&A...397.1097R}.

In the presence of an equatorward meridional flow (with the relative 
direction $\vec{e}_\perp$ perpendicular to the tube axis) the drag 
force, Eq.\ (\ref{fdrag}), tries to push the flux ring to lower
latitudes.
For a stationary mechanical equilibrium to exist (Appdx.\ \ref{equi}),
the component of the drag parallel to the rotation axis, $\vec{e}_z$,
must be balanced by buoyancy, which requires the density contrast 
\begin{equation}
\frac{\rho_0}{\rho_e} 
=
1
-
\frac{C_D}{\pi} 
\frac{v_\perp^2}{a_0}
\frac{ \left( \vec{e}_\perp \cdot \vec{e}_z \right) }{ \left( 
\vec{g}_{eff} \cdot \vec{e}_z \right) }
<
1
\label{dens}
\end{equation}
between the external and internal density, where $\vec{g}_{eff}$ is the
effective gravitational acceleration defined in Eq.\ (\ref{geff}).
The component of the drag force perpendicular to the rotation axis
opposes the magnetic curvature force so that the internal flow
velocity, $v_0$, required to balance the tension force, is 
\begin{equation}
v_0
= 
\Omega R_0
\left( 
\sqrt{ 1 + \frac{c_A^2 - c_D^2}{\left( \Omega R_0 \right)^2} }
-
1
\right)
\ ,
\label{vint}
\end{equation}
where $c_A$ is the Alfv\'en velocity and $c_D$ the modification
caused by the drag force in the presence of meridional circulation,
given in Eq.\ (\ref{c2d}).
The Alfv\'en velocity introduces the magnetic field dependence in the 
equilibrium condition.

\begin{figure}
\includegraphics[width=\hsize]{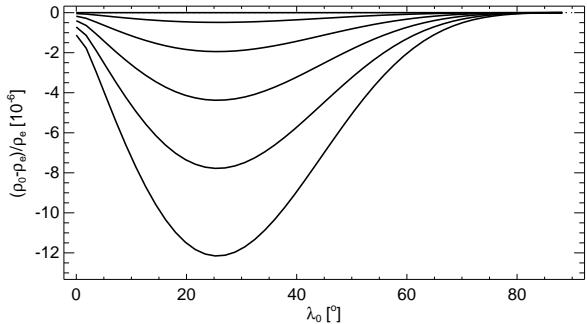}
\caption{Density contrast of a magnetic flux ring in mechanical
equilibrium at latitude $\lambda_0$ and depth $r_0=
5.07\cdot10^{10}\un{cm}$.
The peak velocity of the meridional flow is $v_M= 0, 25, 50, 75, 
100, 125\un{m\cdot s^{-1}}$ (\emph{top to bottom}) and the radius of 
the flux tube $a_0= 10^7\un{cm}$.
}
\label{dens.fig}
\end{figure}

\begin{figure}
\includegraphics[width=\hsize]{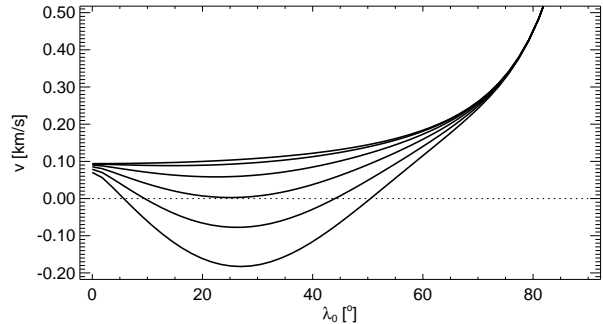}
\caption{Flow velocity inside a magnetic flux ring in mechanical 
equilibrium at latitude $\lambda_0$ and depth $r_0=
5.07\cdot10^{10}\un{cm}$.
The meridional flow velocity is $v_M= 0, 25, 50, 75, 100, 125\un{m\cdot
s^{-1}}$ (\emph{top to bottom}), the radius of the flux tube $a_0=
10^7\un{cm}$, and the field strength $B_0= 15\cdot10^4\un{G}$.
Positive (negative) values correspond to flow velocities faster 
(slower) than the stellar rotation.
}
\label{vint.fig}
\end{figure}

At the assumed equilibrium radius, $r_0= 5.07\cdot10^{10}\un{cm}$, in 
the middle of the overshoot region, the stellar stratification is 
characterised through the pressure scale height $H_p= 
5.52\cdot10^9\un{cm}$, the gravitational acceleration $g=
5.06\cdot10^4\un{cm\cdot s^{-2}}$, the density $\rho_e= 0.154\un{g\cdot
cm^{-3}}$, and the superadiabaticity $\delta= -9.77\cdot10^{-7}$.
For the meridional circulation defined above, the density contrast and
internal flow velocity required for a stationary mechanical equilibrium
are shown in Figs. \ref{dens.fig} and \ref{vint.fig}, respectively.

\subsection{Flux tube simulations}
The stability properties of magnetic flux tubes depend on the stellar 
stratification in its environment (including the meridional flow 
topology) and the magnetic field strength. 
The determination of instability criteria and characteristic growth 
times requires a specific linear stability analysis similar to 
\citet{1982A&A...106...58S, 1988ApJ...333..965V, 1995GAFD...81..233}, 
which is however beyond the scope of the present paper.
Investigations in the case of the Sun and other cool stars consistently
show that the evolution of flux tubes with initial radii larger than
$\sim 1000\un{km}$ are only marginally affected by the drag force, and
that magnetic field strengths $\ga 10^5\un{G}$ are required to initiate
rising flux loops whose properties upon eruption at the surface are in
agreement with observations \citep[e.g.][]{1987ApJ...316..788C,
1994ApJ...436..907F, 1994A&A...281L..69S}.
Here, we consider tubes with initial radii $a_0= 70-1000\un{km}$ and
initial magnetic field strengths $B_0= 10-20\cdot10^4\un{G}$, to
determine the basic influence of these parameters on the latitudinal
eruption pattern in the presence of meridional flows.
The amount of transported magnetic flux, $\Phi= B_0 \pi a_0^2\sim 
10^{19}-6\cdot10^{21}\un{Mx}$, is comparable to that of small pores and
bi-polar spot groups, respectively \citep{2000sostact.book..S}.
We accomplish simulations for the meridional flow pattern defined above
with peak flow velocities $v_M\le 125\un{m\cdot s^{-1}}$.

The simulations start with the perturbation of an equilibrium flux ring
located in the middle of the overshoot region within the range of
latitudes $\lambda_0= 5-75\degr$.
The initial perturbation consists of a superposition of harmonic
displacements with low wave numbers and amplitudes of a few percent of 
the local pressure scale height.
The adiabatic evolution of individual flux tubes is followed using a
non-linear Lagrangian scheme described in more detail in
\citet{1986A&A...166..291M} and \citet{ 1995ApJ...441..886C}.
Owing to the lateral pressure balance, Eq.\ (\ref{pressequi}), the 
summit of a rising flux loop expands as the external pressure
decreases.
The simulations stop just beneath the stellar surface ($r\ga
0.98\un{R_\star}$), where the underlying thin flux tube approximation
becomes inapplicable.
In the upper-most part of the convection zone the rise of the tube 
summit is dominated by strong magnetic buoyancy so that the tube
trajectory is almost radial.
The final point of the simulations thus determines the location of the 
emergence of a bi-polar spot group.

\subsection{Latitudinal distributions of magnetic flux eruption}
\begin{figure}
\includegraphics[width=\hsize]{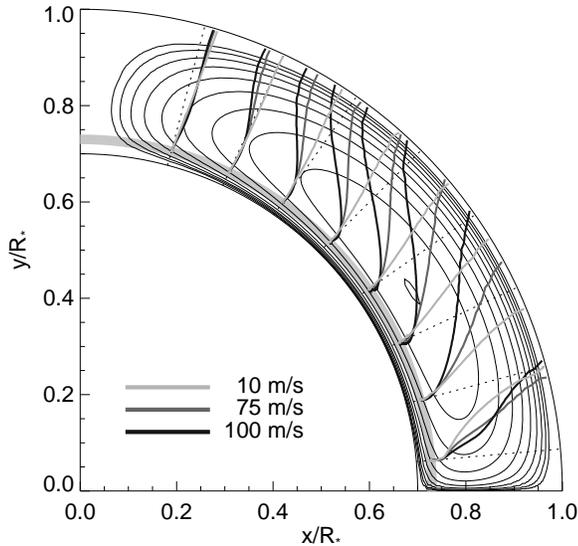}
\caption{
Latitudinal trajectories of the summit of rising flux loops with
initial field strength $B_0= 15\cdot10^4\un{G}$ and tube radius $a_0= 
10^7\un{cm}$ for maximal flow velocities $v_M= 10, 75$, and 
$100\un{m\cdot s^{-1}}$ (\emph{grey shaded lines}).
The flow of the meridional circulation (\emph{solid lines}, $n= 1.5, m=
0.76, k= 0.5, r_b= 0.7\un{R_\star}$) is poleward at the surface and 
equatorward at the bottom of the convection zone.
The shaded region at the bottom of the convection zone marks the 
overshoot region.
}
\label{lattraj.fig}
\end{figure}

\begin{figure}
\includegraphics[width=\hsize]{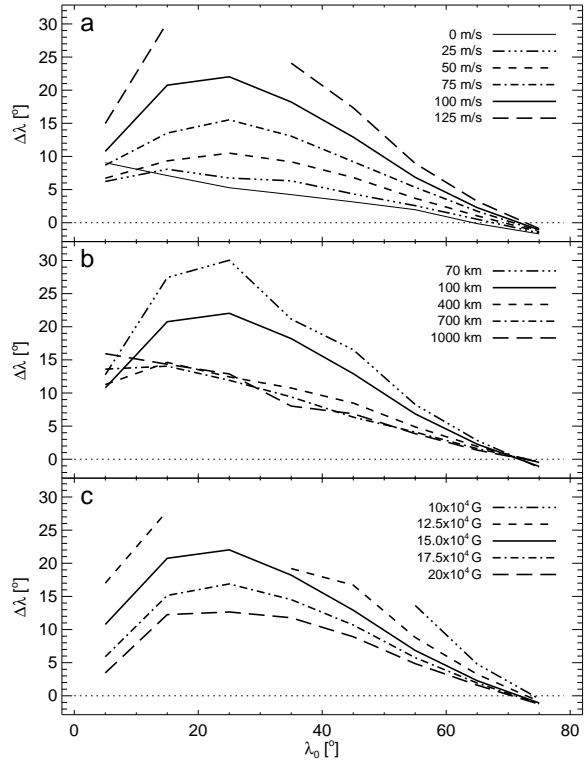}
\caption{
Latitudinal deflection, $\Delta \lambda= \lambda_e - \lambda_0$, of 
erupting flux tubes in the presence of meridional flows with different 
peak flow velocities (Panel \textbf{a}, for $B_0= 15\cdot10^4\un{G}$
and $a_0= 10^7\un{cm}$); initial tube radii (Panel \textbf{b}, for
$v_M= 100\un{m\cdot s^{-1}}$ and $B_0= 15\cdot10^4\un{G}$); initial
magnetic field strengths (Panel \textbf{b}, for $v_M= 100\un{m\cdot
s^{-1}}$ and $a_0= 10^7\un{cm}$).
Gaps in the curves for high flow velocities and low magnetic field
strengths are due to stable flux ring equilibria for the respective set
of parameters.
}
\label{latdist.fig}
\end{figure}

The interplay between Coriolis and magnetic tension force results in a 
deflection of rising flux loops to higher latitudes.
Without a meridional circulation the deflection is less than $\Delta 
\lambda= \lambda_e - \lambda_0\la 10\degr$, with $\lambda_e$ being
the eruption latitude on the stellar surface.
The deflection is greatest for flux tubes which originate from low 
initial latitudes and decreases for higher starting latitudes.
In the presence of meridional flows the poleward deflection is 
considerably larger and shows a characteristic dependence on the 
initial latitude of the flux ring (Fig.\ \ref{lattraj.fig}).
Flux tubes originating from low to intermediate latitudes (here, around
$\lambda_0\sim 25\degr$) are most affected by poloidal flows,
whereas those starting from equatorial or high latitudes are in 
contrast only slightly more deflected than in the absence of an 
external circulation (Fig.\ \ref{latdist.fig}).
This latitude-dependent deflection resembles closely the initial 
density contrast and internal velocity variation (Figs.\ \ref{dens.fig}
and \ref{vint.fig}, respectively), suggesting that the strong 
deflection is partly caused by the equilibrium conditions.

\begin{figure}
\includegraphics[width=\hsize]{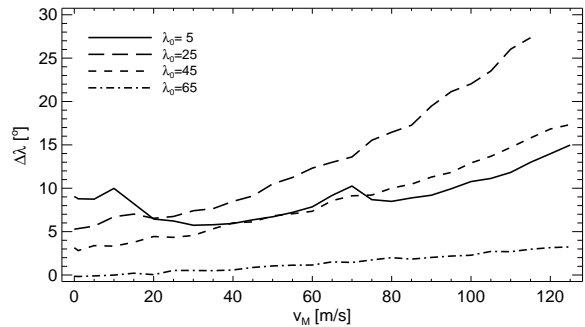}
\caption{
Dependence of the poleward deflection, $\Delta \lambda= \lambda_e - 
\lambda_0$, on the meridional flow velocity, $v_M$.
}
\label{dlamvp.fig}
\end{figure}

The latitudinal distribution patterns obtained for different meridional 
flow velocities are qualitatively similar, with the amplitude of the
pattern increasing with the peak flow velocity (Figs.\
\ref{latdist.fig}a and \ref{dlamvp.fig}).
Whereas flux tubes with smaller radii show a characteristic deflection 
pattern with a maximum at intermediate latitudes, flux tubes with radii 
larger than $200-300\un{km}$ yield an eruption pattern in which the 
poleward deflection continuously decreases from low to high initial 
latitudes (Fig.\ \ref{latdist.fig}b).
Flux tubes with larger initial radii show deflections only slightly 
larger than in the case without a meridional circulation.
For magnetic flux tubes with high field strengths the latitudinal
deflection is weaker (Fig.\ \ref{latdist.fig}c), since the relative
impact of the magnetic buoyancy on the dynamical evolution is larger.
Their trajectory through the convection zone is more radial, with minor
latitudinal deflections only.

\subsection{Comments}
Thin magnetic flux tubes with weak average field strengths are found to
be susceptible to meridional circulations.
In particular flux tubes originating from low to intermediate latitudes
are subject to significant poleward deflections during their rise to 
the stellar surface.
We conjecture that this latitude-, size-, and field strength-depending 
deflection causes the bulk of low-flux elements to erupt at higher 
latitudes than more substantial flux tubes.
Compared to stars with weak or no meridional flows, this separation 
effect causes a bias in the overall surface distribution of erupting 
magnetic flux toward higher latitudes.

The enhanced susceptibility of smaller magnetic flux elements inside
the convection zone to meridional flows is in basic agreement with the 
properties of magnetic flux features observed on the solar surface.
There, smaller flux elements are readily transported to higher 
latitudes, whereas new active regions and sunspots (in their entity)
hardly participate in the poleward motion ($v_M\approx 10\un{m\cdot
s^{-1}}$).
However, after their fragmentation through supergranular motions, the 
dissolving remnants of initially large flux concentrations are swept 
polewards as well.

\section{Stellar butterfly diagrams}
\label{butter}
The latitudinal distribution of erupting flux tubes depends in a 
non-linear way on the magnitude of the meridional circulation as well
as on the magnetic flux and original latitude of the tube inside the
overshoot region.
In the following, we consider the emergence pattern of flux tubes with 
an initial magnetic field strength $B_0= 15\cdot 10^{4}\un{G}$ and an
initial radius $a_0= 100\un{km}$ as representative for the full 
range of parameters considered in Sect.\ \ref{erupt}.

Surface flux transport models require a description of the statistical 
properties of newly emerging bipoles during an activity cycle,
including their emergence rates, latitudes, sizes, fluxes, and tilt
angles.
Since the observational data base on rapidly rotating solar-like stars
is insufficient to create empirical butterfly diagrams, we adopt the
solar butterfly diagram as a qualitative template, which we extrapolate
in latitude and scale according to the results in Sect.\ \ref{erupt}.
The solar butterfly diagram is characterised by emergence latitudes
between about $40\degr$ latitude at the start of a cycle and $5\degr$
at its end.
The width of the wing of each cycle is about $10\degr$ at the beginning
of the cycle and decreases to $5\degr$ at the end.
The cycle period is $11\un{yr}$, excluding one-year overlaps of 
successive cycles \citep{vanBal1998,Mackay2004}.

\subsection{Latitudes of emergence vs. latitudes of origin}
\begin{figure}
\includegraphics[width=\hsize]{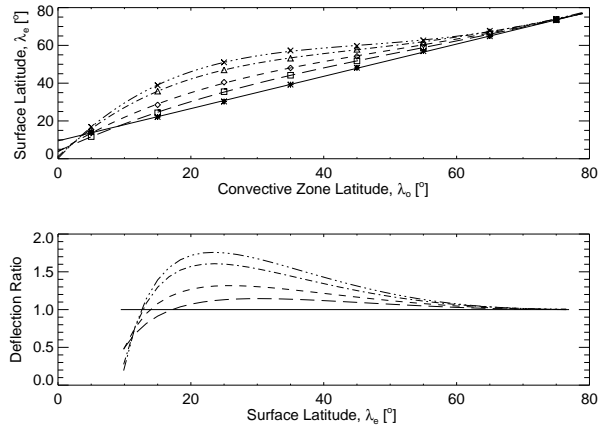}
\caption{Panel \textbf{a}: Latitudes of emergence, $\lambda_e$, of 
magnetic flux tubes as a function of their latitude of origin, 
$\lambda_0$, inside the convection zone.
The initial magnetic field strength of the flux tubes is $B_0= 
15\cdot10^{4}\un{G}$ and the initial radius $a_0= 100\un{km}$.
The flow velocity of the meridional circulation is $v_M= 0\un{m\cdot
s^{-1}}$ (\emph{solid}), $50\un{m\cdot s^{-1}}$ (\emph{long-dashed}),
$75\un{m\cdot s^{-1}}$ (\emph{short-dashed}), $100\un{m\cdot s^{-1}}$
(\emph{dashed-dotted}), and $110\un{m\cdot s^{-1}}$
(\emph{dashed-tripple dotted}); \emph{lines} represent polynomial fits
to the results (\emph{symbols}) of the flux eruption simulations in 
Sect.\ \ref{erupt}.
Panel \textbf{b}:  Deflection ratio, $\lambda_e (v_M) / \lambda_e (0)$,
relative to the case without any meridional circulation.
Values larger (smaller) than one indicate relative deflections toward
the pole (equator).
Note that the abscissa in panel \textbf{b} is equal to the ordinate in
panel \textbf{a}, since the deflection is expressed in terms of
latitudes of emergence.
}
\label{fig:plot1}
\end{figure}

Based on the simulation results in Sect.\ \ref{erupt}, we associate
latitudinal ranges of magnetic flux emergence, $\lambda_e$, with
latitudinal ranges of origin, $\lambda_0$, of the parent flux tubes in
the overshoot region.
For the reference (i.e.\ solar) butterfly diagram, we refer to the case
$v_M= 0\un{m\cdot s^{-1}}$, since the weak solar meridional circulation
with a peak velocity of $\sim 10\un{m\cdot s^{-1}}$ has little effect 
on the sub-surface evolution of rising flux tubes.
In this case, the relationship between $\lambda_e$ and $\lambda_0$ is
virtually linear (Fig.\ \ref{fig:plot1}a, solid line/asterisks), 
and we ascribe the emergence of bipolar regions on the Sun to magnetic 
flux tubes which originate from within the range of latitudes
$\lambda_0\simeq 35-5\degr$.
Further assignments between latitudes of emergence and latitudes of
origin are summarised in Table \ref{lats}.
\begin{table}
\caption{Correspondence between the latitudes of origin, $\lambda_0$, 
and latitudes of emergence, $\lambda_e$, of flux tubes with an initial
field strength of $B_0= 15\cdot10^{4}\un{G}$ and radius $a_0=
100\un{km}$ (see Fig.\ \ref{fig:plot1}).}
\begin{tabular}{cccccc}
\hline
 & & \multicolumn{4}{c}{$\lambda_e [\degr]$} \\
\raisebox{1.5ex}[-1.5ex]{Case} & \raisebox{1.5ex}[-1.5ex]{$\lambda_0
[\degr]$} & $v_M= 0\un{m/s}$ & $50\un{m/s}$ & $75\un{m/s}$ &
$100\un{m/s}$ \\
\hline
S & 35 -- 5 & 40 -- 10 & 44 -- 10 & 47 -- 13 & 54 -- 15 \\
I & 47 -- 3 & 50 -- 10 & 54 -- 8  & 47 -- 9  & 60 -- 9 \\ 
L & 58 -- 3 & 60 -- 10 & 61 -- 9  & 63 -- 9  & 65 -- 9
\end{tabular}
\label{lats}
\end{table}
We shall refer to the small ($\lambda_0= 35-5\degr$), intermediate 
($47-3\degr$), and large ($58-3\degr$) ranges of originating latitudes 
as case S, I, and L, respectively.

\subsection{The wings of the butterfly}
\label{wings}
In contrast to the solar reference case, strong meridional circulations
imply non-linear relationships between the eruption latitudes of
magnetic flux tubes and the latitudes of their origin (Fig.\
\ref{fig:plot1}a) and, consequently, distortions of the wing shapes of
stellar butterfly diagrams.
To transform the solar template into a stellar butterfly diagram, we 
determine the ratio between the latitudes of emergence, $\lambda_e 
(v_M)$, \emph{subject to meridional flows} and the latitudes of 
emergence, $\lambda_e (0)$, of the reference case with \emph{vanishing 
meridional flow}.
The deflection ratios $\lambda_e (v_M) / \lambda_e (0)$ in Fig.\ 
\ref{fig:plot1}b represent mappings, which describe the non-linear 
stretching of the solar template.
In addition to the poleward displacement of flux emergence at
intermediate latitudes also a slight equatorward displacement occurs at
low latitudes.

\begin{figure}
\includegraphics[width=\hsize]{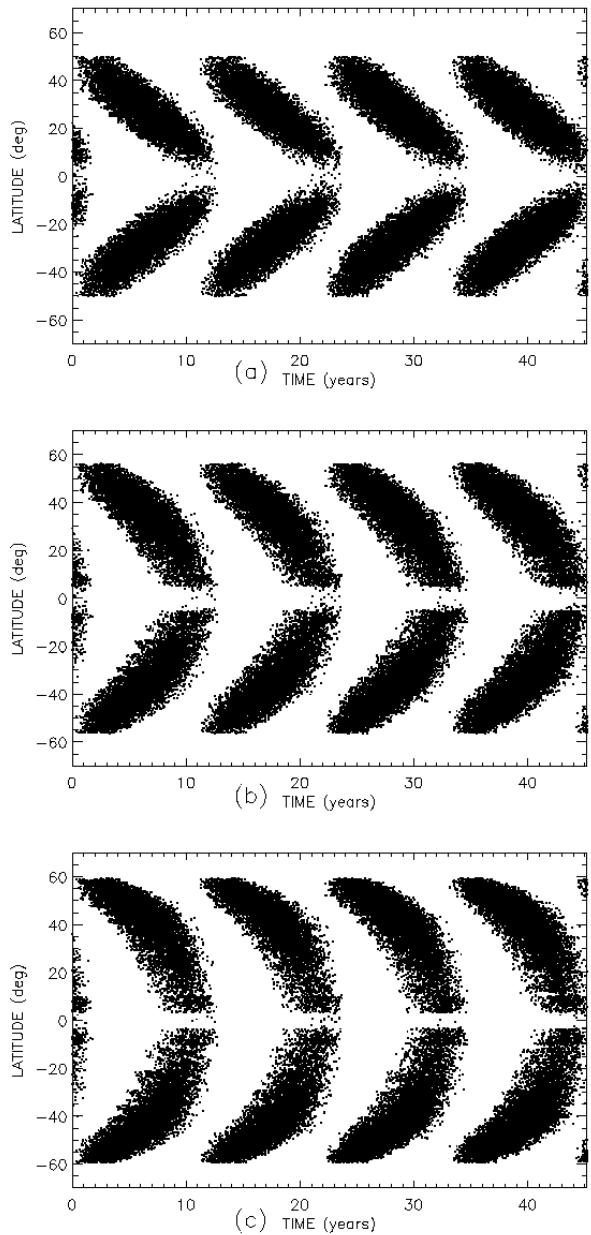}
\caption{Butterfly diagrams over four 11-year activity cycles in the 
presence of meridional circulations with flow velocities $v_M= 
10\un{m\cdot s^{-1}}$ (Panel \textbf{a}); $v_M= 75\un{m\cdot s^{-1}}$ 
(Panel \textbf{b}); $v_M= 100\un{m\cdot s^{-1}}$ (Panel \textbf{c}).
In all cases the generating magnetic flux tubes originate from within 
the range of latitudes $\lambda_0= 47-3\degr$.
The pre-eruptive poleward deflection caused by the meridional 
circulation pushes the upper boundary of the butterfly diagram to higher
latitudes, from $\lambda_e\approx 50\degr$ for the weak flow (Panel
\textbf{a}) to $\lambda_e\approx 60\degr$ for the strong flow (Panel
\textbf{c}).}
\label{fig:plot2} 
\end{figure}

Examples of stellar butterfly diagrams subject to meridional
circulations with different peak velocities are shown in Fig.\
\ref{fig:plot2} (for case I).
Figure \ref{fig:plot2}a depicts a stellar butterfly diagram subject to
a weak meridional circulation (i.e. the solar template linearly
stretched to the latitudinal range $50-10\degr$), whereas the Figs.\
\ref{fig:plot2}b \& c show the result of the non-linear transformation
of the latitudinal pattern of emerging bipoles caused by higher flow
velocities.
The diagrams show that for strong meridional circulations the number of 
bipoles emerging at higher latitudes is increased:  the higher the flow
velocity, the more convex is the temporal evolution of the
stellar butterfly diagram.
Qualitatively similar results are obtained for the cases S and L.

\section{Surface evolution and intermingling of magnetic bipoles}
\label{surft}

\subsection{Basic scenario and empirical constraints}
After the dynamical disconnection of newly emerged bipoles from their
parent flux tubes, the small-scale evolution of magnetic surface
features is governed by local magneto-convective motions and
interactions with ambient magnetic flux concentrations.
Advection of flux with the same (opposite) polarity through the
convective turnover of supergranular cells leads to an enhancement 
(annihilation) of magnetic flux.
Averaged over length scales larger than supergranular cells (in the
case of the Sun about $30\un{Mm}$) the evolution of the magnetic field
resembles a dispersion process \citep{Leighton1964}.
The lifetime of individual magnetic surface features depends on the 
diffusion time scale. 
The evolution of the global magnetic field topology, in turn, depends 
on the relation between the lifetimes of merging flux features and the 
characteristic transport timescales of the differential rotation and 
the meridional circulation.
Differential rotation shears and diverges bipolar spot groups spread
out over a range of latitudes.
The increasing distance between the centres of opposite flux polarities 
reduces the rate of mutual flux annihilation, whereas the associated
decrease of the tilt angle reduces the bias in the latitudinal
distribution of magnetic polarities (see also Appdx.\ \ref{single}).
Merging magnetic surface features are transported to higher latitudes 
by the poleward directed meridional circulation and pile-up in polar 
regions.
Owing to the biased latitudinal polarity distribution, the polar flux 
is dominated by one polarity during half of the magnetic cycle.
With the reversal of the polarity of newly emerging bipolar groups 
after half a magnetic cycle (Hale's law), magnetic flux of the opposite
polarity is predominantly transported toward the pole, which 
annihilates and substitutes the previous flux conglomeration.
This alternating process implies a stable oscillation of the net surface
flux over successive activity cycles, without a persistent accumulation
of magnetic flux.
For a review about the development and limitations of surface flux
transport model see \citet[e.g.][]{2005LRSP....2....5S}.

In contrast to the case of the Sun, where the polar magnetic field is
virtually unipolar, with magnetic flux densities of about
$10\un{Mx\cdot cm^{-2}}$, the surface magnetic fields observed on young
active stars exhibit a high degree of intermingling of magnetic flux
polarities \citep{donati97abdor95, donati99abdor96, donati03} and dark
polar caps \citep{2002AN....323..309S}, indicative for strong magnetic
field strengths.
\citet{Schrijver2001b} found that flux emergence rates thirty times 
larger than on the Sun entail polar flux densities of $300-500\un{Mx 
\cdot cm^{-2}}$, which are deemed to be sufficiently large to suppress
convective upflows of energy, entailing the formation of dark spots
\citep[e.g.][]{1982smhd.book.....P}.
In their simulations, the polar magnetic flux is unipolar and 
surrounded at lower latitudes by a ring of magnetic flux of the 
opposite polarity.
Focusing on the intermingling of polar magnetic flux,
\citet{Mackay2004} identified two key parameters for the generation of
high degrees of polarity mixture: magnetic bipoles have to emerge at
latitudes up to $50-70\degr$, and the peak value of the meridional flow
has to be over $100\un{m\cdot s^{-1}}$.

The simulations of \citet{Mackay2004} disregard the feedback of strong
meridional circulations on the latitudinal pattern of flux eruption.
This feedback is now provided through the consistently determined
stellar butterfly diagrams in Sect.\ \ref{butter}.
In the following, we use the \citeauthor{Mackay2004}-model to 
investigate the impact of the convex wing structures on the polar 
magnetic field properties.

\subsection{Surface flux transport model}
The surface flux transport simulations of \citet{Mackay2004} follow the
radial magnetic field component on the stellar surface regarding the
combined influence of flux emergence, differential rotation, meridional
flow, and supergranular diffusion.
Using spherical harmonics up to degree 63, the spatial resolution is 
about $30\un{Mm}$.
The simulations are based on the extrapolation of solar transport and
flux emerging properties \citep{Gaizauskas1983, Wang1989, Harvey1993,
Schrijver1994, Tain1999} to more rapidly rotating stars \citep[see
also][]{Schrijver2001b}.

The strength and profile of the differential rotation of rapidly
rotating stars is similar to the solar case
\citep[e.g.][]{cameron02diffrot}.
We therefore adopt the solar differential rotation profile given by 
\citet{Snodgrass1983}, which implies the characteristic shear timescale 
$\tau_\Omega= 0.25\un{yr}$.
The meridional flow profile is given by $v_p (\lambda)= - v_M \sin
\left( \pi \lambda / \lambda_p \right)$, whereby the flow velocity
vanishes near the pole above $\lambda_p= 75\degr$
\citep{Hathaway1996}.
The characteristic timescale for the latitudinal flux transport is
$\tau_\mathrm{mf}= R_\odot/v_M$.
Lacking appropriate empirical constraints, it is assumed that 
the supergranular convective profile and turn-over timescales of
rapidly rotating stars are solar-like, implying a diffusion coefficient
of $D= 450\un{km^2\cdot s^{-1}}$ \citep{vanBal1998} and a diffusion
time scale for magnetic features with length scale $l$ of $\tau_D=
l^2/D$ ($= 34\un{yr}$ for $l= R_\odot$).

Following \citet{Schrijver2001b}, we take the net amount of
magnetic flux emerging during each cycle to be thirty times the solar
value, that is $3.15\cdot10^{26}\un{Mx}$.
New flux is injected into the evolving surface magnetic field 
distribution at random longitudes in the form of 13\,200 bipolar 
magnetic regions, whose average tilt angles vary as $\lambda_e/2$
\citep{Wang1989}.
In each successive cycle the polarity of the preceding and following
flux of the bipoles alternates according to Hale's law.
The simulations cover a time span of four activity cycles to verify 
that stable oscillations of the polar field from one cycle to the next
are obtained.
For further details about model assumptions, input parameters, and 
initial conditions see \citet[ Sect.\ 2--4]{Mackay2004}.

In the following, we focus on the total (unsigned) magnetic flux,
\begin{equation}
\Phi_\mathrm{tot} (t)
=
\int | B_r (R,\theta,\phi,t) | d S
\ ,
\label{phitotdef}
\end{equation}
the total (unsigned) \emph{polar} magnetic flux in each
hemisphere,
\begin{equation}
\Phi_\mathrm{[np,sp]} (t)
=
\int | B_r (R,\theta_{[\theta<20\degr,\theta>160\degr]},\phi,t) | d S
\ ,
\label{phipolardef}
\end{equation}
the imbalance of positive and negative magnetic flux in the polar
regions,
\begin{equation}
\delta_\mathrm{[n,s]} (t)
=
\int B_r (R,\theta_{[\theta<20\degr,\theta>160\degr]},\phi,t) d S
\ ,
\label{phiinbaldef}
\end{equation}
and the polar magnetic flux in the \emph{northern} hemisphere,
\begin{equation}
\Phi_{[+,-]} (t)
=
\int B_r (R,\theta< 20\degr,\phi,t)_{[B_r>0,B_r<0]} d S
\ .
\label{phinorthdef}
\end{equation}
The principal quantities of the analysis will be mean magnetic flux
densities, that is the respective magnetic flux divided by the surface
over which it is calculated \citep[see][]{Mackay2004}.
The last quantity, Eq.\ (\ref{phinorthdef}) is a measure for the degree
of intermingling of opposite polarity elements at high latitudes.

\subsection{Surface magnetic field properties of rapid rotators}
\label{simus}
\begin{figure*} 
\includegraphics[width=\hsize]{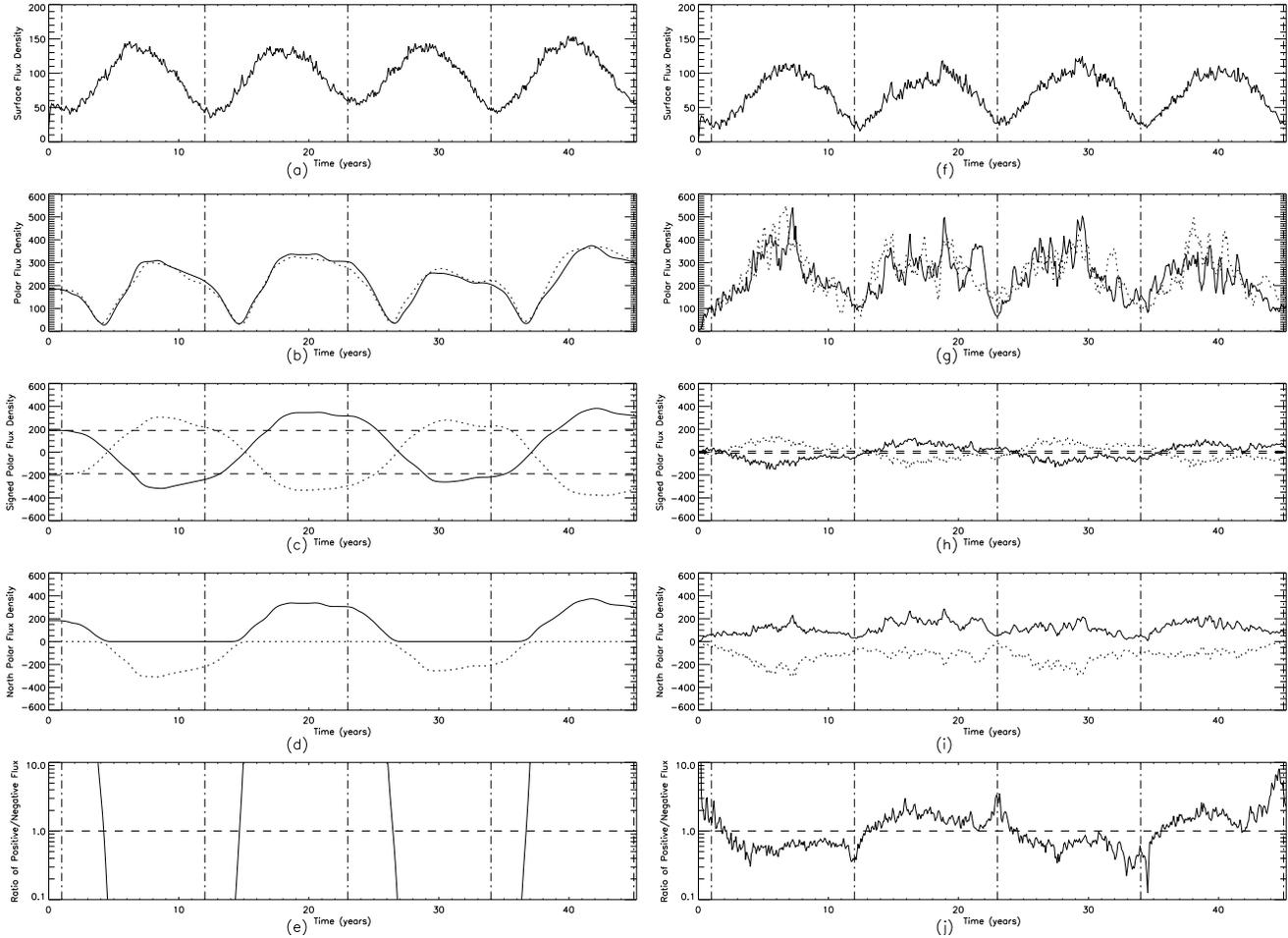}
\caption{Results of the surface flux transport simulations for 
meridional circulations with a weak flow velocity of $11\un{m\cdot 
s^{-1}}$ (Panels \textbf{a}--\textbf{e}), and a strong flow velocity of
$100\un{m\cdot s^{-1}}$ (Panels \textbf{f}--\textbf{j}); the associated
butterfly diagrams are shown in Figs.\ \ref{fig:plot2}a and
\ref{fig:plot2}c, respectively, with generating flux tubes originating
from within the range of latitudes $\lambda_0= 47-3\degr$ (Case I).
Panels \textbf{a} \& \textbf{e}: mean surface flux density, $\Phi_{tot}
/(4\pi R^2)$; panels \textbf{b} \& \textbf{g}: mean polar flux density 
of the northern (\emph{solid}) and southern (\emph{dotted}) 
hemispheres; panels \textbf{c} \& \textbf{h}: signed polar flux 
densities in the northern (\emph{solid}) and southern (\emph{dotted}) 
hemisphere, with \emph{dashed} lines representing the initial values; 
panels \textbf{d} \& \textbf{i}: positive (\emph{solid}) and negative 
(\emph{dotted}) polar flux density of the northern hemisphere; panels 
\textbf{e} \& \textbf{j}: ratio between the positive and negative polar
flux density of the northern hemisphere.
Each quantity is determined once every 27 days.
\emph{Dashed-dotted} lines mark the start of a new cycle. 
All flux densities are in $\un{Mx\cdot cm^{-2}}$.  
}
\label{fig:plot3}
\end{figure*}

We consider weak and strong meridional flows of the case I to detail
our results (Fig.\ \ref{fig:plot3}).
For both flow velocities the mean surface flux density, 
$\Phi_\mathrm{tot}/(4\pi R^2)$, is in phase with the activity cycle
emergence rate (Fig.\ \ref{fig:plot3}a \& f).
If the meridional flow velocity is high, then more magnetic flux is
pushed to higher latitudes, where the proximity of opposite bipole
polarities increases the flux annihilation rate.
In this case, the mean surface flux densities are consequently smaller
than for slow meridional circulations.
The resulting variation of the mean \emph{polar} flux density (i.e.\
beyond $70\degr$ latitude) is shown in Figs.\ \ref{fig:plot3}b \& g.
In the case of weak meridional flows this quantity shows a noticeable
phase shift relative to the mean surface flux density, with a
\emph{cycle-averaged} polar flux density of $\sim 192\un{G}$.
In the case of fast flows the polar field rises and falls in phase,
with a significantly higher cycle-averaged polar flux density of $\sim
270\un{G}$.
Breaking up the polar flux density imbalance (Figs.\ \ref{fig:plot3}c 
\& h) according to the contributions of positive and negative flux 
(Fig.\ \ref{fig:plot3}d \& i) shows that for slow meridional flow 
velocities the polar field is unipolar, with a field reversal midway
through the cycle.
For fast flow velocities, in contrast, both magnetic polarities are 
present throughout the cycle and rise and fall in strength together.
The ratio between the positive and negative flux densities in the polar
caps is thus in both cases different (Fig.\ \ref{fig:plot3}e \& j).
For weak flows there is hardly any intermingling of flux of opposite
polarities at high latitudes, except for a limited time span during
field reversal, when little magnetic flux is located within the polar 
cap.
For strong meridional flows, in contrast, the intermingling of magnetic
flux with opposite polarities is about $65\%$, averaged over an
activity cycle.

\begin{figure*}
\includegraphics[width=\hsize]{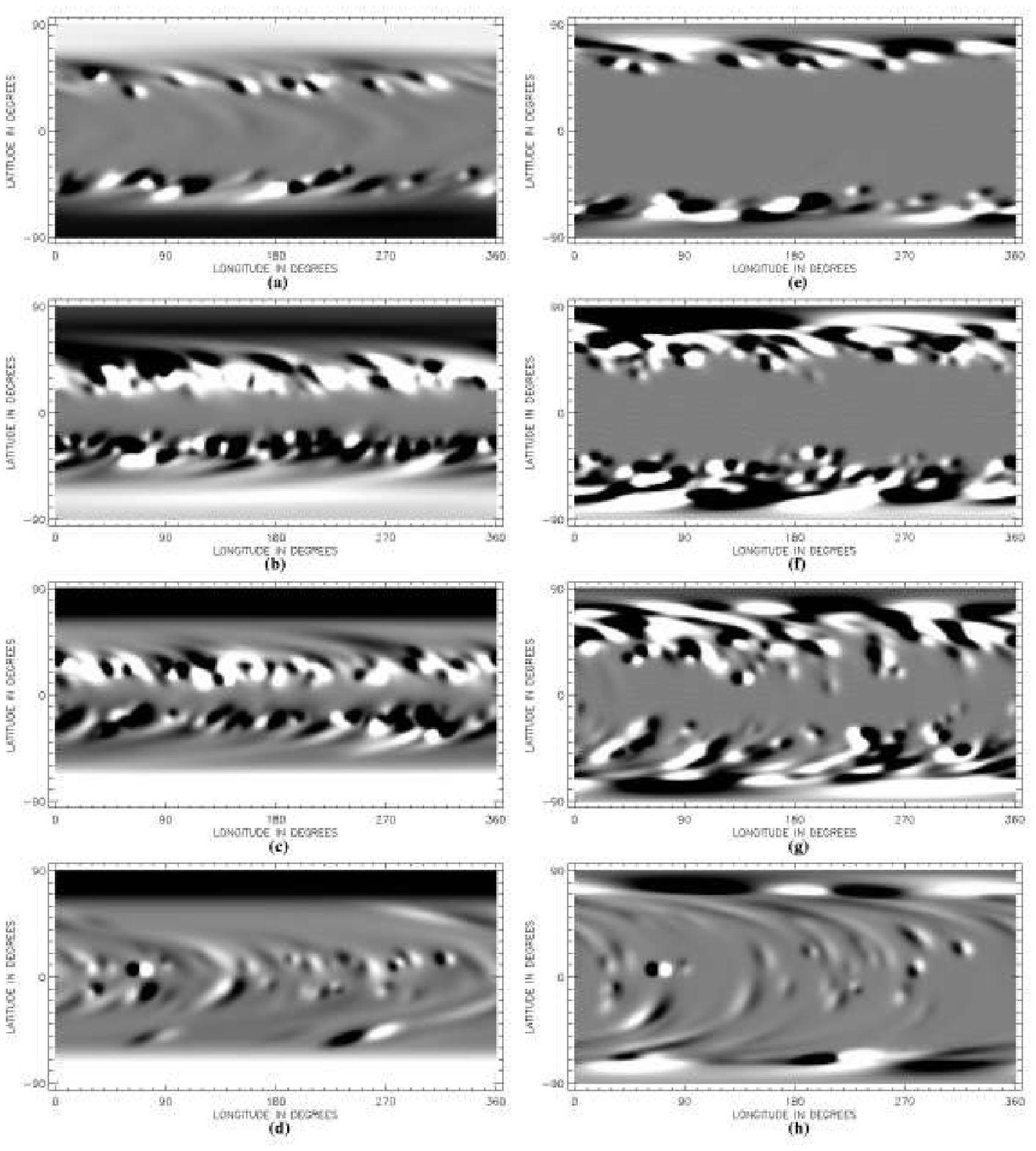}
\caption{Maps of the radial magnetic field component in the case of
weak ($10\un{m\cdot s^{-1}}$, left column) and strong ($100\un{m\cdot 
s^{-1}}$, right column) meridional flow velocities.
The associated butterfly diagrams (cf.\ Fig.\ \ref{fig:plot2}a \& c)
are generated by magnetic flux tubes which originate from within the
latitudinal range $\lambda_0= 47-3\degr$.
Each set shows snapshots after year 1 (cycle minimum, Panels \textbf{a}
\& \textbf{e}), 5 (before maximum, Panels \textbf{b} \& \textbf{f}), 8 
(after maximum, Panels \textbf{c} \& \textbf{g}), and 11 (declining 
phase, Panels \textbf{d} \& \textbf{h}).
\emph{White} and \emph{black} shadings represent positive and negative
magnetic flux densities, respectively, saturating at $\pm
300\un{Mx\cdot cm^{-2}}$.
}
\label{fig:plot6}
\end{figure*}

Snapshots of maps showing the radial magnetic field component (Fig.\ 
\ref{fig:plot6}) illustrate that in the case of weak flows the magnetic
flux in polar regions is dominated by one polarity only, whereas in the
case of fast flows the field is intermingled.

\subsubsection{Parameter study}
\begin{figure}
\includegraphics[width=\hsize]{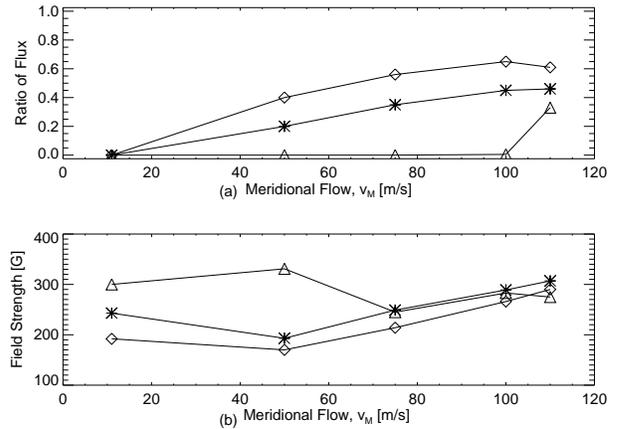}
\caption{Panel \textbf{a}: Cycle-averaged ratio between the positive 
and negative magnetic flux within the northern polar cap as a function 
of the meridional flow velocity, $v_M$, for originating latitudes 
$\lambda_0= 35-5\degr$ (\emph{triangles}); $47-3\degr$
(\emph{diamonds}); and $57-3\degr$ (\emph{asterisks}).
Panel \textbf{b}: Cycle-averaged polar flux densities.
}
\label{fig:plot4} 
\end{figure}

For the latitudinal ranges I and L, the degree of intermingling of
opposite polarities within the polar regions exceeds $20\%$ for
meridional flow velocities beyond about $40\un{m\cdot s^{-1}}$ (Fig.\
\ref{fig:plot4}a).
The fact that in case I the intermingling is higher than in case L
indicates that merely increasing both the meridional flow velocity and
the latitudinal range of flux origin does not produce per se higher
degrees of intermingling.
Instead, there exists an optimal combination of latitudes of origin and
meridional flow velocities for which the intermingling within polar
caps peaks.

The associated polar flux densities are on the order of 
$200-300\un{Mx\cdot cm^{-2}}$ (Fig. \ref{fig:plot4}b), with values
increasing (decreasing) with the meridional flow velocity above (below)
$v_M\sim 50\un{m\cdot s^{-1}}$.
This behaviour is due to the different transport- and annihilation- 
timescales (see Appdx.\ \ref{single}).
In the case of strong meridional flows, the advection of magnetic flux
to higher latitudes is more efficient than the decrease caused by flux
annihilation as the magnetic surface features are transported
poleward.
The same argument holds for the increase of the polar flux density with
larger latitudinal ranges of flux origin, since for bipoles emerging at 
higher latitudes the shorter migration time towards the pole implies
less flux annihilation.
For meridional flow velocities around $100\un{m\cdot s^{-1}}$ the polar
flux densities become sufficiently large ($\sim 300\un{G}$) to suppress
convection.

In case S, the polar magnetic flux is unipolar throughout the activity
cycle, without a significant intermingling of opposite polarities.
The average polar flux densities of $\approx 300\un{G}$ are almost 
independent of the meridional flow velocity.

In summary, we find that a combination of intermediate ranges of
originating latitudes (here, $\lambda_0 = 47-3\degr$) with fast
meridional flow velocities ($v_M\sim 100\un{m\cdot s^{-1}}$) is optimal
for the production of both strong intermingling of polarities and high
field strengths in polar regions.
This suggests that in rapidly rotating solar-like stars magnetic flux
originates from latitudinal ranges somewhat larger than in the case of
the Sun.

\subsubsection{Effect of the pre-eruptive poleward deflection} 
\citet{Mackay2004} disregard the influence of meridional circulations
on the sub-surface evolution of magnetic flux.
In their investigation the reference (i.e.\ solar) butterfly diagram is
linearly stretched to different latitudinal ranges, leaving the
structure of the wings unaltered.
According to our results in Sect.\ \ref{wings}, the eruption pattern
\emph{within} a given range of emergence latitudes depends on the 
meridional flow velocity, with more bipoles emerging at higher 
latitudes the faster the flow velocity.
We compare the two approaches in the case of a meridional circulation 
with a flow velocity of $100\un{m\cdot s^{-1}}$.
Keeping other model parameters unchanged, including the pre-eruptive
poleward deflection entails about $20\%$ higher polar flux densities
than in the original approach of \citeauthor{Mackay2004}
This result applies to all ranges of generating latitudes  (Table
\ref{comp}).
\begin{table}
\caption{Comparison of the cycle-averaged polar magnetic field
strength, $\bar{B}$, including and excluding the effect of pre-eruptive
deflection of magnetic flux tubes for a meridional circulation with
flow velocity $v_M= 100\un{m\cdot s ^{-1}}$.
The latter values are taken from \protect\citet{Mackay2004}.}
\begin{tabular}{ccc}
\hline
 & $\bar{B}\ [G]$ & $\bar{B}\ [G]$ \\
\raisebox{1.5ex}[-1.5ex]{$\lambda_e\ [\degr]$} & 
incl.\ defl. & excl.\ defl. \\
\hline
50 -- 10 & 248 & 200 \\
60 -- 10 & 266 & 218 \\
65 -- 10 & 290 & 236
\end{tabular}
\label{comp}
\end{table}
The reason for this increase lies in the more convex structure of the
wings of the butterfly (cf.\ Fig.\ \ref{fig:plot2}).
The larger number of emerging bipoles at higher latitudes over an 
activity cycle implies shorter transport times (on average) toward the 
pole during which less flux annihilation occurs.

\section{Discussion}
\label{disc}
The present work relates observed properties of magnetic flux on the
stellar surface to the sub-surface evolution of the originating
magnetic fields and, therefore, to the underlying dynamo processes
inside the convective envelope.
Our results extend the findings of \citet{Mackay2004}, who identified 
high latitudes of magnetic flux emergence and high meridional flow 
velocities as the key ingredients for a strong mixing of opposite 
polarities in polar regions of rapidly rotating stars.
We find that the enhanced pre-eruptive poleward deflection of rising 
flux tubes inside the convection zone provides a consistent explanation
for the required larger latitudinal range for newly emerging bipoles.

Within the framework of our model assumptions, a solar-like butterfly 
diagram with flux emergence up to $40\degr$ latitudes is generated by 
erupting magnetic flux tubes, which originate at the bottom of the
convection zone from latitudes $\la 35\degr$.
Owing to the slight prolateness of the solar tachocline 
\citep[e.g.][]{1999ApJ...527..445C}, these low latitudes coincide in 
the case of the Sun well with the region of strong shear flows, 
efficient magnetic field amplification, and field storage.
For more rapidly rotating stars, the butterfly structures producing the
optimal levels of polar flux intermingling are generated by erupting
flux tubes, which originate from higher latitudes, $\la 50\degr$, as
well.
This suggest the existence of a larger latitudinal range of efficient
dynamo operation than in the case of the Sun, implying a possibly less
prolate tachocline.

Based on the results of \citet{Schrijver2001b}, we assumed a magnetic
flux emergence rate thirty times larger than the solar value to produce
polar flux densities sufficiently strong to entail dark caps.
Dynamo operation is generally expected to be more efficient at higher 
rotation rates, mainly entailing an increase of the surface filling 
factor of magnetic features \citep[e.g.][]{1991LNP...380..389S}.
The larger latitudinal range of dynamo operation suggested by our 
results complies with this expectation of a larger amount of magnetic 
flux generated during an activity cycle.
Considering a possible concomitant increase of the average magnetic 
field strength, we note that strong-field flux tubes are less 
susceptible to meridional circulations and therefore hamper the 
formation of high flux densities and high degrees of flux intermingling
in polar regions.
We also find that flux tubes with initial radii $a_0\ga 1000\un{km}$,
which are expected to lead to the formation of average sunspots, are
only marginally affected by meridional flows.
Since the magnitude of the drag force scales $\propto v_\perp^2 / a$, a
meridional flow velocity about three times faster would correspond 
roughly to a flux tube radius about ten times smaller, entailing a 
noticeable change of the eruption latitude.
In view of the uncertainties of the sub-surface flow profile, this
estimate indicates that also larger magnetic flux elements could be
affected by meridional circulations.

By using the solar butterfly diagram as an empirical template, we
circumvent the treatment of the magnetic field amplification and
dissolving mechanisms inside the stellar convection zone, which as yet
lack a consistent theoretical description.
This approach, however, implies that the activity signatures of rapidly
rotating stars are due to a solar-like dynamo.
The assumption of an $\alpha\Omega$-dynamo mechanism has important
implications, for example, on the period, and even the mere existence, 
of the activity cycle.
The $11\un{yr}$ activity cycle assumed here sets the time scale for the
pile-up of magnetic flux in polar regions.
If we, for example, increase the cycle length, then we would obtain
qualitatively similar results by increasing the total amount of 
magnetic flux emerging per cycle.
A conclusive parameter study of different ratios between eruption, 
transport, and diffusion time scales is however beyond the scope of 
this work; a discussion of the impact of selected surface transport
model parameters is given by \citet{Mackay2004}.

Our work supplements earlier investigations, which considered the 
evolution of the weak solar magnetic field subject to a prescribed 
meridional flow pattern and a dynamo wave at the bottom of the 
convection zone \citep[e.g][]{1994A&A...291..975D, 1995A&A...303L..29C,
1999SoPh..184...61C}.
All approaches, however, suffer from the same lack of knowledge about
the sub-surface profiles of meridional circulation and their dependence
on the rotation, mass, structure, and evolutionary stage of the star.
Owing to the density difference between the top and the bottom of the 
convection zone, the equatorward flow is slower than the surface flow,
and the larger the radial extent of the equatorward flow region, the 
lower the associated velocities will be.
The latitudinal flow velocity at the bottom of the convection zone
possibly has an influence on the propagation velocity of dynamo waves
and, consequently, on the period of the activity cycle.
But in absence of an appropriate theory, this impact cannot be
quantified.
The circulation model applied here assumes a ratio of 2:1 between the 
upper convection zone flowing poleward and the lower convection zone 
flowing toward the equator.
A different radial profile would not only affect the equator- and
poleward drift of rising flux tubes on their way to the surface, but
also their equilibrium and stability properties inside the overshoot
region.
The former aspect has to be investigated in a parameter study 
considering both meridional flow and differential rotation profiles as 
well as different stellar rotation periods.
Furthermore, specifically determined tilt angles and eruption time 
scales of newly emerging bipoles have to be consistently included in 
the surface flux transport model.

Compared with a mere linear stretching of the solar butterfly diagram
to higher latitudes \citep[e.g.][]{Mackay2004}, the additional
convexity of the butterfly wings promotes the intermingling of polar
flux even more, which eases the necessity for high poleward flow
velocities.
Although the meridional flow velocities of $\sim 100\un{m\cdot s^{-1}}$
are lower than the values suggested by \citet{Mackay2004}, they are yet
still about ten times larger than in the case of the Sun.
Whereas in the case of the Sun, with $v_M\approx 11\un{m\cdot s^{-1}}$,
the bias in the latitudinal size distribution of newly emerging bipoles
is probably neither discernible nor relevant, convex butterfly 
structures and the size distributions of emerging dipoles may become 
observable on rapidly rotating stars once the temporal and spatial 
resolutions are sufficiently high.
Strong meridional circulations on the order of $100\un{m\cdot s^{-1}}$
are also expected to cause a shift of the stellar eigenfrequencies, in 
addition to the splitting of eigenfrequencies caused by differential 
rotation (M Roth, priv.\ comm.)
Depending on the sub-surface profile of the flow, a meridional 
circulation of $100\un{m\cdot s^{-1}}$ would cause in the case of the
Sun a frequency shift of a few tenth of $\mu$Hz.
Albeit small, this effect would be observable on the Sun.
Given sufficiently long and complete time sequences, astroseismological
observations may provide constraints on the magnitude (and possible
structure) of stellar meridional flows.

\section{Conclusion}
\label{conc}
Strong meridional circulations enhance the poleward deflection of
rising magnetic flux tubes inside stellar convection zones.
Flux tubes which comprise small amounts of magnetic flux and which
originate between low to intermediate latitudes within the overshoot 
region are particularly susceptible to meridional flows.
The resulting latitudinal distribution of newly emerging bipoles 
renders the wing structure of stellar butterfly diagrams distinctively
more convex than in the solar case.
The larger amount of magnetic flux emerging at higher latitudes
supports the formation of high magnetic flux densities and a
significant intermingling of opposite-polarity magnetic flux in polar
regions, which is in agreement with recent observational results of
rapidly rotating solar-like stars.

The strong pre-eruptive deflection of magnetic flux tubes provides a
consistent explanation for the required high latitudes of flux
emergence identified by previous investigations \citep{Mackay2004}.
The synergetic combination of pre-eruptive flux tube deflection and 
post-eruptive bipole transport yields higher values for the average
polar flux densities than in the previous approach, and thus eases the
necessity for high meridional flow velocities.
The functional dependence of the polar field properties on the extent
and structure of the stellar butterfly diagram makes it possible to
conjecture about potential regions of efficient dynamo operation at the
bottom of the convection zone.
In rapidly rotating cool stars, we suppose the latitudinal range of 
magnetic flux production to extend to higher latitudes ($\la 50\degr$) 
than in the case of the Sun ($\la 35\degr$).

\section*{Acknowledgements}
The authors would like to thank M Roth and M Sch\"ussler for helpful
comments and F Moreno-Insertis, P Caligari, and A van Ballegooijen, who
initially developed the codes used in this paper.
VH and DHM acknowledge financial support for their research through 
PPARC grands.

\bibliographystyle{mn2e}
\bibliography{mf1aph}

\begin{thebibliography}{}

\bibitem[\protect\citeauthoryear{{Baliunas}, {Donahue}, {Soon} \& {and 17
  co-authors}}{{Baliunas} et~al.}{1995}]{1995ApJ...438..269B}
{Baliunas} S.~L.,  {Donahue} R.~A.,  {Soon} W.~H.,    {and 17 co-authors} 1995,
  ApJ, 438, 269

\bibitem[\protect\citeauthoryear{{Basu} \& {Antia}}{{Basu} \&
  {Antia}}{2001}]{2001MNRAS.324..498B}
{Basu} S.,  {Antia} H.~M.,  2001, Mon. Not. Royal Astron. Soc., 324, 498

\bibitem[\protect\citeauthoryear{{Baumann}, {Schmitt}, {Sch{\"u}ssler} \&
  {Solanki}}{{Baumann} et~al.}{2004}]{2004A&A...426.1075B}
{Baumann} I.,  {Schmitt} D.,  {Sch{\"u}ssler} M.,    {Solanki} S.~K.,  2004,
  A\&A, 426, 1075

\bibitem[\protect\citeauthoryear{{Berdyugina}}{{Berdyugina}}{2004}]{2004SoPh..%
224..123B}
{Berdyugina} S.~V.,  2004, Sol. Phys., 224, 123

\bibitem[\protect\citeauthoryear{{Caligari}, {Moreno-Insertis} \&
  {Sch\"ussler}}{{Caligari} et~al.}{1995}]{1995ApJ...441..886C}
{Caligari} P.,  {Moreno-Insertis} F.,    {Sch\"ussler} M.,  1995, ApJ, 441, 886

\bibitem[\protect\citeauthoryear{{Caligari}, {Sch\"ussler} \&
  {Moreno-Insertis}}{{Caligari} et~al.}{1998}]{1998ApJ...502..481C}
{Caligari} P.,  {Sch\"ussler} M.,    {Moreno-Insertis} F.,  1998, ApJ, 502, 481

\bibitem[\protect\citeauthoryear{{Charbonneau}, {Christensen-Dalsgaard},
  {Henning}, {Larsen}, {Schou}, {Thompson} \& {Tomczyk}}{{Charbonneau}
  et~al.}{1999}]{1999ApJ...527..445C}
{Charbonneau} P.,  {Christensen-Dalsgaard} J.,  {Henning} R.,  {Larsen} R.~M.,
  {Schou} J.,  {Thompson} M.~J.,    {Tomczyk} S.,  1999, ApJ, 527, 445

\bibitem[\protect\citeauthoryear{{Choudhuri}}{{Choudhuri}}{1989}]{1989SoPh..12%
3..217C}
{Choudhuri} A.~R.,  1989, Sol. Phys., 123, 217

\bibitem[\protect\citeauthoryear{{Choudhuri} \& {Dikpati}}{{Choudhuri} \&
  {Dikpati}}{1999}]{1999SoPh..184...61C}
{Choudhuri} A.~R.,  {Dikpati} M.,  1999, Sol. Phys., 184, 61

\bibitem[\protect\citeauthoryear{{Choudhuri} \& {Gilman}}{{Choudhuri} \&
  {Gilman}}{1987}]{1987ApJ...316..788C}
{Choudhuri} A.~R.,  {Gilman} P.~A.,  1987, ApJ, 316, 788

\bibitem[\protect\citeauthoryear{{Choudhuri}, {Schussler} \&
  {Dikpati}}{{Choudhuri} et~al.}{1995}]{1995A&A...303L..29C}
{Choudhuri} A.~R.,  {Schussler} M.,    {Dikpati} M.,  1995, A\&A, 303, L29+

\bibitem[\protect\citeauthoryear{{Collier Cameron}}{{Collier
  Cameron}}{2001}]{2001astr.conf..183C}
{Collier Cameron} A.,  2001, LNP Vol.~573: Astrotomography, Indirect Imaging
  Methods in Observational Astronomy, pp 183--206

\bibitem[\protect\citeauthoryear{{Collier Cameron}, {Donati} \&
  {Semel}}{{Collier Cameron} et~al.}{2002}]{cameron02diffrot}
{Collier Cameron} A.,  {Donati} J.-F.,    {Semel} M.,  2002, Mon. Not. Royal
  Astron. Soc., 330, 699

\bibitem[\protect\citeauthoryear{DeVore, Sheeley \& Boris}{DeVore
  et~al.}{1984}]{DeVore1984}
DeVore C.,  Sheeley N.,    Boris J.,  1984, Sol. Phys., 21, 1

\bibitem[\protect\citeauthoryear{{Dikpati} \& {Choudhuri}}{{Dikpati} \&
  {Choudhuri}}{1994}]{1994A&A...291..975D}
{Dikpati} M.,  {Choudhuri} A.~R.,  1994, A\&A, 291, 975

\bibitem[\protect\citeauthoryear{{Donahue}}{{Donahue}}{1996}]{Donahue1996}
{Donahue} R.~A.,  1996, in IAU Symp. 176: Stellar Surface Structure {Long-term
  stellar activity: three decades of observations}.
p.~261

\bibitem[\protect\citeauthoryear{{Donati} \& {Brown}}{{Donati} \&
  {Brown}}{1997}]{donati97}
{Donati} J.-F.,  {Brown} S.~F.,  1997, A\&A, 326, 1135

\bibitem[\protect\citeauthoryear{{Donati}, {Cameron}, {Semel}, {Hussain},
  {Petit}, {Carter}, {Marsden}, {Mengel}, {L{\' o}pez Ariste}, {Jeffers} \&
  {Rees}}{{Donati} et~al.}{2003}]{2003MNRAS.345.1145D}
{Donati} J.-F.,  {Cameron} A.~C.,  {Semel} M.,  {Hussain} G.~A.~J.,  {Petit}
  P.,  {Carter} B.~D.,  {Marsden} S.~C.,  {Mengel} M.,  {L{\' o}pez Ariste} A.,
   {Jeffers} S.~V.,    {Rees} D.~E.,  2003, Mon. Not. Royal Astron. Soc., 345,
  1145

\bibitem[\protect\citeauthoryear{{Donati}, {Cameron}, {Semel}, {Hussain},
  {Petit}, {Carter}, {Marsden}, {Mengel}, {L{\'o}pez Ariste}, {Jeffers} \&
  {Rees}}{{Donati} et~al.}{2003}]{donati03}
{Donati} J.-F.,  {Cameron} A.~C.,  {Semel} M.,  {Hussain} G.~A.~J.,  {Petit}
  P.,  {Carter} B.~D.,  {Marsden} S.~C.,  {Mengel} M.,  {L{\'o}pez Ariste} A.,
  {Jeffers} S.~V.,    {Rees} D.~E.,  2003, Mon. Not. Royal Astron. Soc., 345,
  1145

\bibitem[\protect\citeauthoryear{{Donati} \& {Collier Cameron}}{{Donati} \&
  {Collier Cameron}}{1997}]{donati97abdor95}
{Donati} J.-F.,  {Collier Cameron} A.,  1997, Mon. Not. Royal Astron. Soc.,
  291, 1

\bibitem[\protect\citeauthoryear{{Donati}, {Collier Cameron}, {Hussain} \&
  {Semel}}{{Donati} et~al.}{1999}]{donati99abdor96}
{Donati} J.-F.,  {Collier Cameron} A.,  {Hussain} G.~A.~J.,    {Semel} M.,
  1999, Mon. Not. Royal Astron. Soc., 302, 437

\bibitem[\protect\citeauthoryear{{D'Silva} \& {Choudhuri}}{{D'Silva} \&
  {Choudhuri}}{1993}]{1993A&A...272..621D}
{D'Silva} S.,  {Choudhuri} A.~R.,  1993, A\&A, 272, 621

\bibitem[\protect\citeauthoryear{{Fan}, {Fisher} \& {McClymont}}{{Fan}
  et~al.}{1994}]{1994ApJ...436..907F}
{Fan} Y.,  {Fisher} G.~H.,    {McClymont} A.~N.,  1994, ApJ, 436, 907

\bibitem[\protect\citeauthoryear{{Ferriz-Mas} \& {Sch\"ussler}}{{Ferriz-Mas} \&
  {Sch\"ussler}}{1993}]{1993GAFD...72..209}
{Ferriz-Mas} A.,  {Sch\"ussler} M.,  1993, Geophys. Astrophys. Fluid Dyn., 72,
  209

\bibitem[\protect\citeauthoryear{{Ferriz-Mas} \& {Sch\"ussler}}{{Ferriz-Mas} \&
  {Sch\"ussler}}{1995}]{1995GAFD...81..233}
{Ferriz-Mas} A.,  {Sch\"ussler} M.,  1995, Geophys. Astrophys. Fluid Dyn., 81,
  233

\bibitem[\protect\citeauthoryear{{Gaizauskas}, {Harvey}, {Harvey} \&
  {Zwaan}}{{Gaizauskas} et~al.}{1983}]{Gaizauskas1983}
{Gaizauskas} V.,  {Harvey} K.~L.,  {Harvey} J.~W.,    {Zwaan} C.,  1983, ApJ,
  265, 1056

\bibitem[\protect\citeauthoryear{{Granzer}, {Sch\"ussler}, {Caligari} \&
  {Strassmeier}}{{Granzer} et~al.}{2000}]{2000A&A...355.1087G}
{Granzer} T.,  {Sch\"ussler} M.,  {Caligari} P.,    {Strassmeier} K.~G.,  2000,
  A\&A, 355, 1087

\bibitem[\protect\citeauthoryear{{Harvey} \& {Zwaan}}{{Harvey} \&
  {Zwaan}}{1993}]{Harvey1993}
{Harvey} K.~L.,  {Zwaan} C.,  1993, Sol. Phys., 148, 85

\bibitem[\protect\citeauthoryear{{Hathaway}}{{Hathaway}}{1996}]{Hathaway1996}
{Hathaway} D.~H.,  1996, ApJ, 460, 1027

\bibitem[\protect\citeauthoryear{{Holzwarth}}{{Holzwarth}}{2004}]{2004AN....32%
5..408H}
{Holzwarth} V.,  2004, Astron. Nachr., 325, 408

\bibitem[\protect\citeauthoryear{{Holzwarth} \& {Sch{\" u}ssler}}{{Holzwarth}
  \& {Sch{\" u}ssler}}{2001}]{2001A&A...377..251H}
{Holzwarth} V.,  {Sch{\" u}ssler} M.,  2001, A\&A, 377, 251

\bibitem[\protect\citeauthoryear{{Holzwarth} \& {Sch{\" u}ssler}}{{Holzwarth}
  \& {Sch{\" u}ssler}}{2003}]{2003A&A...405..303H}
{Holzwarth} V.,  {Sch{\" u}ssler} M.,  2003, A\&A, 405, 303

\bibitem[\protect\citeauthoryear{{Jeffers}, {Donati} \& {Cameron}}{{Jeffers}
  et~al.}{2005}]{jeffers2005abdor}
{Jeffers} S.,  {Donati} J.,    {Cameron} A.,  2005, Mon. Not. Royal Astron.
  Soc.

\bibitem[\protect\citeauthoryear{{Jetsu}, {Tuominen}, {Grankin}, {Mel'Nikov} \&
  {Schevchenko}}{{Jetsu} et~al.}{1994}]{1994A&A...282L...9J}
{Jetsu} L.,  {Tuominen} I.,  {Grankin} K.~N.,  {Mel'Nikov} S.~Y.,
  {Schevchenko} V.~S.,  1994, A\&A, 282, L9

\bibitem[\protect\citeauthoryear{{Korhonen} \& {Elstner}}{{Korhonen} \&
  {Elstner}}{2005}]{2005A&A...440.1161K}
{Korhonen} H.,  {Elstner} D.,  2005, A\&A, 440, 1161

\bibitem[\protect\citeauthoryear{{Leighton}}{{Leighton}}{1964}]{Leighton1964}
{Leighton} R.~B.,  1964, ApJ, 140, 1547

\bibitem[\protect\citeauthoryear{{Mackay}, {Jardine}, {Cameron}, {Donati} \&
  {Hussain}}{{Mackay} et~al.}{2004}]{Mackay2004}
{Mackay} D.~H.,  {Jardine} M.,  {Cameron} A.~C.,  {Donati} J.-F.,    {Hussain}
  G.~A.~J.,  2004, Mon. Not. Royal Astron. Soc., 354, 737

\bibitem[\protect\citeauthoryear{{Mackay} \& {Lockwood}}{{Mackay} \&
  {Lockwood}}{2002}]{Mackay2002b}
{Mackay} D.~H.,  {Lockwood} M.,  2002, Sol. Phys., 209, 287

\bibitem[\protect\citeauthoryear{{Moreno-Insertis}}{{Moreno-Insertis}}{1983}]{%
1983A&A...122..241M}
{Moreno-Insertis} F.,  1983, A\&A, 122, 241

\bibitem[\protect\citeauthoryear{{Moreno-Insertis}}{{Moreno-Insertis}}{1986}]{%
1986A&A...166..291M}
{Moreno-Insertis} F.,  1986, A\&A, 166, 291

\bibitem[\protect\citeauthoryear{{Moreno-Insertis}, {Sch\"ussler} \&
  {Ferriz-Mas}}{{Moreno-Insertis} et~al.}{1992}]{1992A&A...264..686M}
{Moreno-Insertis} F.,  {Sch\"ussler} M.,    {Ferriz-Mas} A.,  1992, A\&A, 264,
  686

\bibitem[\protect\citeauthoryear{{Moss}}{{Moss}}{2004}]{2004MNRAS.352L..17M}
{Moss} D.,  2004, Mon. Not. Royal Astron. Soc., 352, L17

\bibitem[\protect\citeauthoryear{{Ossendrijver}}{{Ossendrijver}}{2003}]{2003A&%
ARv..11..287O}
{Ossendrijver} M.,  2003, A\&AR, 11, 287

\bibitem[\protect\citeauthoryear{{Parker}}{{Parker}}{1966}]{1966ApJ...145..811%
P}
{Parker} E.~N.,  1966, ApJ, 145, 811

\bibitem[\protect\citeauthoryear{{Parker}}{{Parker}}{1975}]{1975ApJ...198..205%
P}
{Parker} E.~N.,  1975, ApJ, 198, 205

\bibitem[\protect\citeauthoryear{{Priest}}{{Priest}}{1982}]{1982smhd.book.....%
P}
{Priest} E.~R.,  1982, Solar Magneto-Hydrodynamics.
Geophysics and Astrophysics Monographs, D. Reidel

\bibitem[\protect\citeauthoryear{{Rempel}}{{Rempel}}{2003}]{2003A&A...397.1097%
R}
{Rempel} M.,  2003, A\&A, 397, 1097

\bibitem[\protect\citeauthoryear{{Saar}}{{Saar}}{1991}]{1991LNP...380..389S}
{Saar} S.~H.,  1991, in {Tuominen} I.,  {Moss} D.,   {R\"udiger} G.,  eds, {The
  Sun and cool stars: activity, magnetism, dynamos, IAU Coll. 130} Lecture
  Notes in Physics, vol. 380, {Recent advances in the observation and analysis
  of stellar magnetic fields}.
Springer-Verlag, pp 389--400

\bibitem[\protect\citeauthoryear{{Schrijver} \& {Harvey}}{{Schrijver} \&
  {Harvey}}{1994}]{Schrijver1994}
{Schrijver} C.~J.,  {Harvey} K.~L.,  1994, Sol. Phys., 150, 1

\bibitem[\protect\citeauthoryear{{Schrijver} \& {Title}}{{Schrijver} \&
  {Title}}{1999}]{1999SoPh..188..331S}
{Schrijver} C.~J.,  {Title} A.~M.,  1999, Sol. Phys., 188, 331

\bibitem[\protect\citeauthoryear{{Schrijver} \& {Title}}{{Schrijver} \&
  {Title}}{2001}]{Schrijver2001b}
{Schrijver} C.~J.,  {Title} A.~M.,  2001, ApJ, 551, 1099

\bibitem[\protect\citeauthoryear{{Schrijver} \& {Zwaan}}{{Schrijver} \&
  {Zwaan}}{2000}]{2000sostact.book..S}
{Schrijver} C.~J.,  {Zwaan} C.,  2000, {Solar and Stellar Magnetic Activity}.
Cambride Astrophysics Series, Cambridge University Press

\bibitem[\protect\citeauthoryear{{Sch{\"u}ssler}}{{Sch{\"u}ssler}}{2005}]{2005%
AN....326..194S}
{Sch{\"u}ssler} M.,  2005, Astron. Nachr., 326, 194

\bibitem[\protect\citeauthoryear{{Sch\"ussler}, {Caligari}, {Ferriz-Mas} \&
  {Moreno-Insertis}}{{Sch\"ussler} et~al.}{1994}]{1994A&A...281L..69S}
{Sch\"ussler} M.,  {Caligari} P.,  {Ferriz-Mas} A.,    {Moreno-Insertis} F.,
  1994, A\&A, 281, L69

\bibitem[\protect\citeauthoryear{{Sch\"ussler}, {Caligari}, {Ferriz-Mas},
  {Solanki} \& {Stix}}{{Sch\"ussler} et~al.}{1996}]{1996A&A...314..503S}
{Sch\"ussler} M.,  {Caligari} P.,  {Ferriz-Mas} A.,  {Solanki} S.~K.,    {Stix}
  M.,  1996, A\&A, 314, 503

\bibitem[\protect\citeauthoryear{{Sch{\"u}ssler} \& {Rempel}}{{Sch{\"u}ssler}
  \& {Rempel}}{2005}]{2005A&A...441..337S}
{Sch{\"u}ssler} M.,  {Rempel} M.,  2005, A\&A, 441, 337

\bibitem[\protect\citeauthoryear{{Sch\"ussler} \& {Solanki}}{{Sch\"ussler} \&
  {Solanki}}{1992}]{1992A&A...264L..13S}
{Sch\"ussler} M.,  {Solanki} S.~K.,  1992, A\&A, 264, L13

\bibitem[\protect\citeauthoryear{{Semel}}{{Semel}}{1989}]{semel89}
{Semel} M.,  1989, A\&A, 225, 456

\bibitem[\protect\citeauthoryear{{Sheeley}}{{Sheeley}}{2005}]{2005LRSP....2...%
.5S}
{Sheeley} N.~R.,  2005, Living Reviews in Solar Physics, 2, 5

\bibitem[\protect\citeauthoryear{{Snodgrass}}{{Snodgrass}}{1983}]{Snodgrass198%
3}
{Snodgrass} H.~B.,  1983, ApJ, 270, 288

\bibitem[\protect\citeauthoryear{{Spruit}}{{Spruit}}{1981}]{1981A&A...102..129%
S}
{Spruit} H.~C.,  1981, A\&A, 102, 129

\bibitem[\protect\citeauthoryear{{Spruit} \& {van Ballegooijen}}{{Spruit} \&
  {van Ballegooijen}}{1982}]{1982A&A...106...58S}
{Spruit} H.~C.,  {van Ballegooijen} A.~A.,  1982, A\&A, 106, 58

\bibitem[\protect\citeauthoryear{{Stix}}{{Stix}}{1989}]{1989sun.book.....S}
{Stix} M.,  1989, The Sun -- An Introduction.
Astronomy \& Astrophysics Library, Springer Verlag

\bibitem[\protect\citeauthoryear{{Strassmeier}}{{Strassmeier}}{2002}]{2002AN..%
..323..309S}
{Strassmeier} K.~G.,  2002, Astron. Nachr., 323, 309

\bibitem[\protect\citeauthoryear{{Tian}, {Zhang}, {Tong} \& {Jing}}{{Tian}
  et~al.}{1999}]{Tain1999}
{Tian} L.,  {Zhang} H.,  {Tong} Y.,    {Jing} H.,  1999, Sol. Phys., 189, 305

\bibitem[\protect\citeauthoryear{{Tobias}, {Brummell}, {Clune} \&
  {Toomre}}{{Tobias} et~al.}{2001}]{2001ApJ...549.1183T}
{Tobias} S.~M.,  {Brummell} N.~H.,  {Clune} T.~L.,    {Toomre} J.,  2001, ApJ,
  549, 1183

\bibitem[\protect\citeauthoryear{{van Ballegooijen}}{{van
  Ballegooijen}}{1982}]{1982A&A...113...99V}
{van Ballegooijen} A.~A.,  1982, A\&A, 113, 99

\bibitem[\protect\citeauthoryear{{van Ballegooijen}, {Cartledge} \&
  {Priest}}{{van Ballegooijen} et~al.}{1998}]{vanBal1998}
{van Ballegooijen} A.~A.,  {Cartledge} N.~P.,    {Priest} E.~R.,  1998, ApJ,
  501, 866

\bibitem[\protect\citeauthoryear{{van Ballegooijen} \& {Choudhuri}}{{van
  Ballegooijen} \& {Choudhuri}}{1988}]{1988ApJ...333..965V}
{van Ballegooijen} A.~A.,  {Choudhuri} A.~R.,  1988, ApJ, 333, 965

\bibitem[\protect\citeauthoryear{{Wang}, {Nash} \& {Sheeley}}{{Wang}
  et~al.}{1989}]{Wang1989a}
{Wang} Y.-M.,  {Nash} A.~G.,    {Sheeley} N.~R.,  1989, Science, 245, 712

\bibitem[\protect\citeauthoryear{{Wang} \& {Sheeley}}{{Wang} \&
  {Sheeley}}{1989}]{Wang1989}
{Wang} Y.-M.,  {Sheeley} N.~R.,  1989, Sol. Phys., 124, 81

\end{thebibliography}

\appendix

\section{Meridional flow model}
\label{meriflow}
\begin{figure}
\includegraphics[width=\hsize]{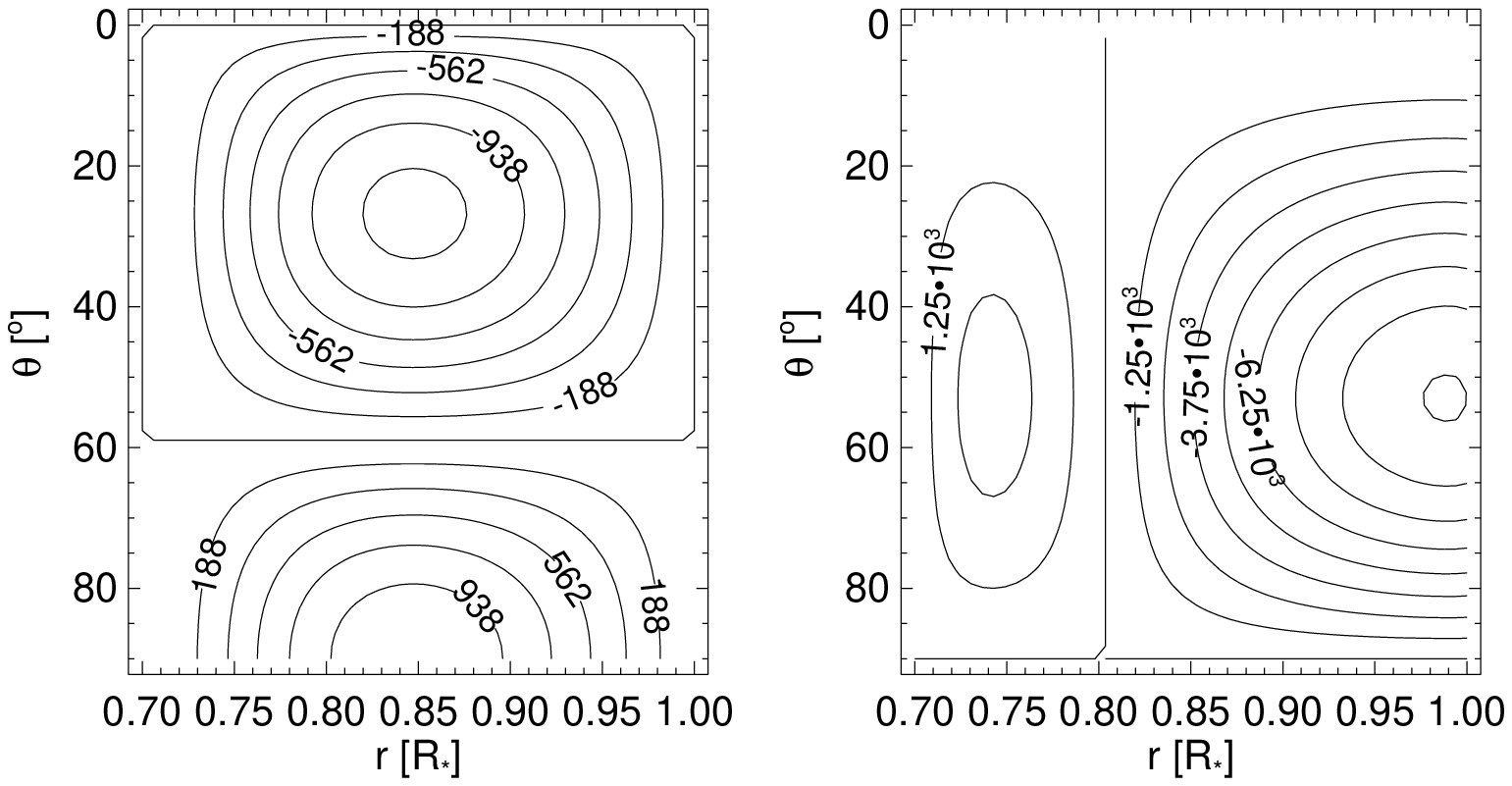}
\caption{
Radial (\emph{left}) and latitudinal (\emph{right}) velocity components
of the meridional flow for the parameter $n= 1.5, m= 0.76, k= 0.5$, 
and $r_b= 0.7 R_\star$.
The maximum of the latitudinal flow velocity on the stellar surface is
located at $\theta_M= 53^o$.
At the radius $\sim 0.8 R_\star$, the latitudinal flow pattern changes
its sign from poleward (above) to equatorward (below).
}
\label{profile.fig}
\end{figure}

We use the analytical model of \citet{1988ApJ...333..965V} for a 
parametrised description of the meridional circulation inside the
convection zone \citep[see also][]{1994A&A...291..975D}.
The poloidal flow pattern, $\vec{v}_p = \left( \nabla \times \Psi
\vec{e}_\phi \right) / \rho_e$, is expressed in terms of the scalar
function, $\Psi$, to ensure that the stationary continuity equation is 
consistently fulfilled.
Under the assumption of a separation of variables, the stream function
\begin{equation}
\Psi \left(r,\theta\right) r \sin \theta
=
u_0 R_\star^2 \rho_e (r) \cdot F(r) \cdot G \left(\theta\right)
\ ,
\label{stream}
\end{equation}
is expressed in terms of the univariate functions 
\begin{eqnarray}
F (r)
& = &
\left( 
 - 
 \frac{1}{n+1} 
 + 
 \frac{c_1}{2n+1} \xi^n 
 - 
 \frac{c_2}{2n + k + 1} \xi^{n+k} 
\right) 
\xi 
\label{fxi}
\\
G (\theta)
& = &
\sin^{m+2} \theta \cos \theta
\ ,
\label{gtheta}
\end{eqnarray}
with the rescaled radial coordinate
\begin{equation}
\xi = \frac{R_\star}{r} - 1
\ .
\label{xi}
\end{equation}
The coefficients
\begin{eqnarray}
c_1
& = & 
\frac{ \left( 2n+1 \right) \left( n+k \right)}{\left( n+1 \right) k}
\xi_b^{-n}
\label{c1coeff}
\\
c_2
& = &
\frac{ \left( 2n+k+1 \right) n}{\left( n+1 \right) k}
\xi_b^{-\left(n+k\right)}
\label{c2coeff}
\end{eqnarray}
depend on the location of the lower boundary, $\xi_b= R_\star / r_b -
1$, of the flow pattern, with $r_b$ being a free model parameter.
The poloidal components of the flow velocity are
\begin{eqnarray}
v_{p,r} 
& = &
u_0 
\left( \frac{R_\star}{r} \right)^2
\frac{F}{\sin \theta} 
\frac{\partial}{\partial \theta} G
\label{vprappdx}
\\
v_{p,\theta} 
& = &
-
u_0 R_\star^2
\frac{G}{r \sin \theta} 
\left( 
 \frac{\partial}{\partial r} F + F \frac{d \ln \rho_e}{d r} 
\right)
\\
& = &
-
u_0
\frac{G}{\sin \theta} 
\left( 1 + \xi \right)^3
\left[
 \left( 1 - \frac{n-\mathcal{H} (\xi)}{n+1} \right)
\right.
\nonumber \\ & & {}
\left.
 -
 c_1 \left( 1 - \frac{n-\mathcal{H} (\xi)}{2n+1} \right) \xi^n
\right.
\nonumber \\ & & {}
\left.
 +
 c_2 \left( 1 - \frac{n-\mathcal{H} (\xi)}{2n+k+1} \right) \xi^{n+k} 
\right]
\label{vpthetappdx}
\end{eqnarray}
with 
\begin{equation}
\mathcal{H}
= 
- \frac{r^2 \xi}{R_\star} \frac{d \ln \rho_e}{d r}
= 
\frac{r^2 \xi}{R_\star H_\rho}
\ ,
\label{densstrat}
\end{equation}
and the local density scale height $H_\rho= - \left( d \ln \rho_e / d r
\right)$.
If the density stratification is approximated through the expression
$\rho_e (r)= C \xi^n$, then it is $\mathcal{H}= n$ and Eq.\
(\ref{vpthetappdx}) results in Eq.\ (\ref{vptheta}).
For an adiabatically stratified convection zone with a ratio of 
specific heats $\gamma= 5/3$, it is $n= 1.5$
\citep{1988ApJ...333..965V}.
The maximum of the latitudinal flow velocity is located at co-latitude
$\theta_M$, specified by the condition $\tan \theta_M = \sqrt{m + 1}$.
We use the value $m= 0.76$ to obtain a solar-like value $\theta_M\simeq
53\degr$.
The flow pattern is composed of a downflow at high latitudes (i.e.\ 
$v_{p,r}< 0$ for $0< \theta< 58\degr$) and an upflow at low latitudes 
(i.e.\ $v_{p,r}> 0$ for $58\degr< \theta< 90\degr$).
The latitudinal velocity component changes its direction at about $0.8
R_\star$, that is $v_{p,\theta}> 0$ (equatorward) for $0.7\le r/R_\star 
\la 0.8$ and $v_{p,\theta}< 0$ (poleward) for $0.8\la 
r/R_\star \le 1$ (Fig.\ \ref{profile.fig}).
At the surface, $r= R_\star$ it is $\xi= 0$ and $v_{p,\theta} = - u_0
\sin^{m+1} \theta \cos \theta$.
The amplitude,
\begin{equation}
u_0
=
-
\frac{v_M}{\sin^{m+1} \theta_M \cos \theta_M}
\label{defu0}
\end{equation}
of the circulation is determined through the maximal flow velocity, 
$v_M$.

\section{Flux tube equilibrium}
\label{equi}
The mechanical equilibrium in the presence of meridional flows is 
determined in the frame of reference co-rotating with the star.
For a homogeneous flux tube (i.e.\ vanishing derivatives with respect 
to the arc length), the stationary equation of motion is
\begin{equation}
 \rho 
 \left( v^2 - \frac{B^2}{4\pi \rho} \right)
 \frac{1}{R}
 \vec{n} 
 =
-
\nabla \left( p + \frac{B^2}{8\pi} \right)
+
\rho
\vec{g}_{eff}
+
2
\rho
\left( \vec{v} \times \vec{\Omega} \right)
\ ,
\label{eqmb}
\end{equation}
where $\vec{v}$ is the internal flow velocity, $\vec{B}$ the magnetic 
field strength, $\rho$ the density, $p$ the gas pressure, $R$ the local
radius of curvature, $\vec{n}$ the normal vector of the tube,
$\vec{\Omega}= \Omega \vec{e}_z$ the stellar rotation vector, and
\begin{equation}
\vec{g}_{eff}
=
\vec{g} 
- 
\vec{\Omega} \times \left( \vec{\Omega} \times \vec{r} \right)
=
g (r)
\left(
 \vec{e}_r
 +
 \vec{e}_R \frac{\Omega^2 r}{g} \sin \theta
\right)
\label{geff}
\end{equation}
the effective gravitational acceleration comprising the centrifugal
contribution, which (for the model assumptions specified in Sect.
\ref{fteq}) inside the overshoot region is of the order $\mathcal{O} 
\left( \Omega^2 r / g \right) \sim 10^{-4}$.
Anticipating the meridional circulation to be of minor importance for
the stellar structure, the external stratification is approximately
hydrostatic,
\begin{equation}
\nabla p_e
=
\rho_e
\vec{g}_{eff}
\ .
\label{eqmwob}
\end{equation}
Using the lateral pressure balance, Eq.\ (\ref{pressequi}), the 
difference between Eqs. (\ref{eqmb}) and (\ref{eqmwob}) yields
\begin{equation}
\left( v^2 - \frac{B^2}{4\pi \rho} \right) \kappa \vec{n} 
=
\left( 1 - \frac{\rho_e}{\rho} \right)
\vec{g}_{eff}
+
2
\left( \vec{v} \times \vec{\Omega} \right)
+
\vec{f}_D
\ ,
\label{eqmft}
\end{equation}
with the hydrodynamic drag force, $\vec{f}_D$, accounting for the 
influence of the distorted external flow on the dynamics of the flux 
tube.
For toroidal magnetic flux tubes with a constant radius of curvature,
the component of Eq.\ (\ref{eqmft}) tangential to the tube axis (i.e.\
the azimuthal component) vanishes per se.
The bi-normal component (i.e.\ parallel to axis of stellar rotation,
$\vec{e}_z$) implies the condition
\begin{equation}
\left( 1 - \frac{\rho_e}{\rho} \right)
\left( \vec{g}_{eff} \cdot \vec{e}_z \right)
=
-
\frac{1}{\rho}
\left( \vec{f}_D \cdot \vec{e}_z \right)
\ .
\label{paracomp}
\end{equation}
With the definition for $\vec{f}_D$ in Eq.\ (\ref{fdrag}) this yields 
the density contrast given in Eq.\ (\ref{dens}), which is required to 
balance the influence of the drag force along the z-axis through 
buoyancy.
The component perpendicular to the stellar rotation axis ($\vec{e}_R=
-\vec{n}$), 
\begin{equation}
\left( v^2 - \frac{B^2}{4\pi \rho} \right) \frac{1}{R}
+
\left( 1 - \frac{\rho_e}{\rho} \right)
\left( \vec{g}_{eff} \cdot \vec{e}_R \right)
+
2 v \Omega
=
-
\frac{1}{\rho}
\left( \vec{f}_D \cdot \vec{e}_R \right)
\label{perpcomp}
\end{equation}
shows that a meridional flow toward the equator in the overshoot region
supports the outward directed inertia and Coriolis force in balancing
the magnetic tension force of the curved flux ring, where $R= r \cos
\lambda$ is the constant radius of curvature.
From the alternative form of Eq.\ (\ref{perpcomp}),
\begin{equation}
\frac{v^2}{R}
+
2 v \Omega
+
\frac{c_D^2}{R}
=
\frac{c_A^2}{R}
\label{velo}
\end{equation}
follows the internal flow velocity given in Eq.\ (\ref{vint}), which is
required to keep the flux ring in mechanical equilibrium.
The contribution by the drag force is formally expressed in terms of 
the velocity
\begin{equation}
\frac{c_D^2}{R_0}
=
\left( 1 - \frac{\rho_e}{\rho_0} \right)
\left( \vec{g}_{eff} \cdot \vec{e}_R \right)
+
\frac{\rho_e}{\rho_0}
\frac{C_D}{\pi} 
\frac{v_\perp^2}{a_0}
\left( \vec{e}_\perp \cdot \vec{e}_R \right)
\label{c2d}
\end{equation}
whereas the Alfv\'en velocity $c_A^2= B^2/(4\pi\rho)$ contains the
dependence on the magnetic field strength.

The particular susceptibility of magnetic flux rings located at low to 
intermediate latitudes (cf.\ Figs.\ \ref{dens.fig} \& \ref{vint.fig}) 
is likely a characteristic feature of the equilibrium properties.
Neglecting centrifugal forces, the latitudinal variation of the density
contrast, Eq.\ (\ref{dens}), is proportional to $v_\perp^2 \cot
\lambda$.
Albeit the actual meridional flow profile inside the stars is unknown,
the increase and decrease of the latitudinal flow velocity from the
pole down the equator is to yield a peak at intermediate latitudes.
In combination with the factor $\cot \lambda$, which continuously
increases toward the equator, the peak of the density contrast will be
located at low to mid latitudes.

\section{Evolution of single bipoles} 
\label{single}
\begin{figure}
\includegraphics[width=\hsize]{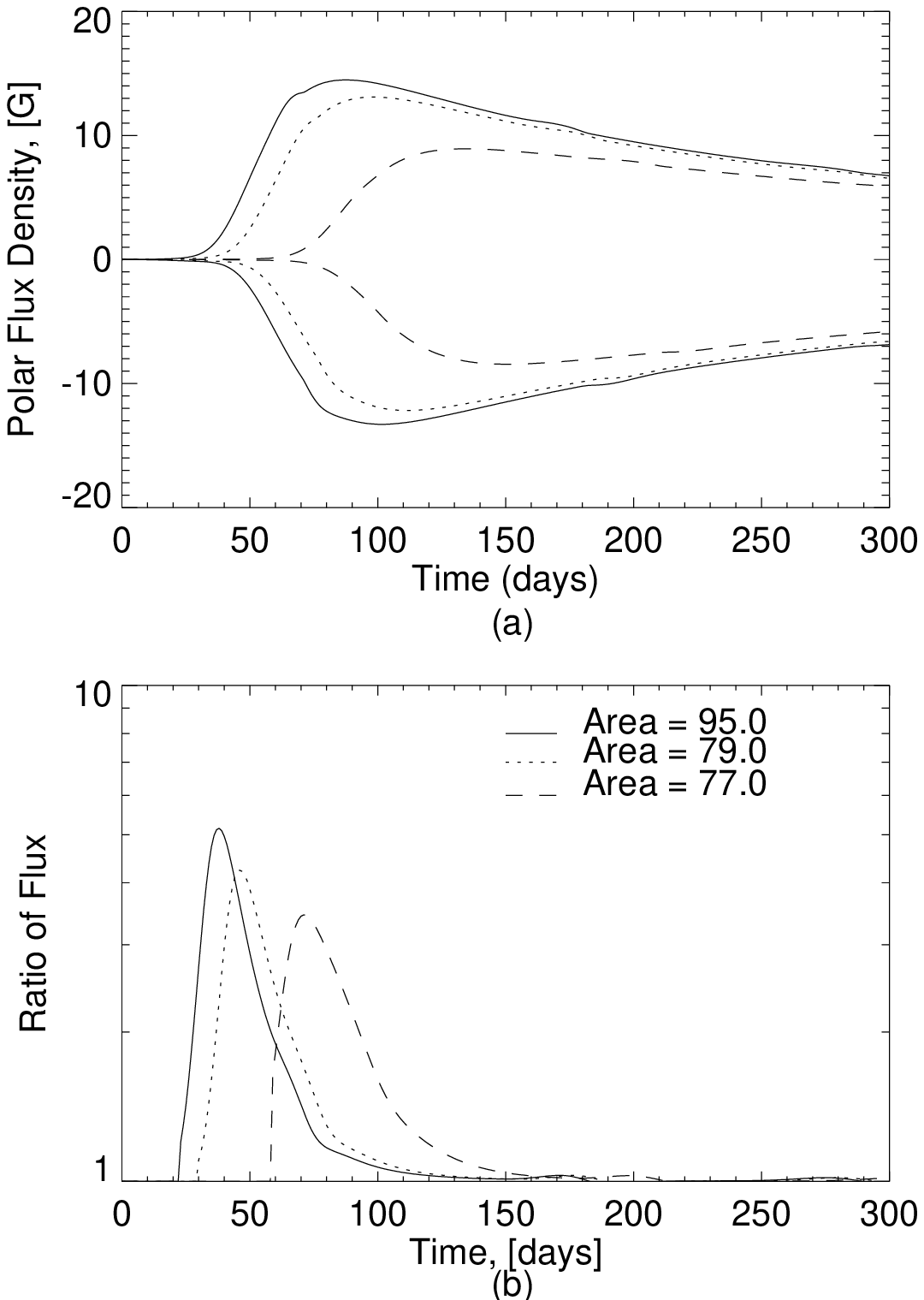}
\caption{Panel \textbf{a}: North polar flux density for a single bipole
inserted at a mean latitude of emergence, $\bar{\lambda}= 21\degr$
(\emph{dashed}), $39\degr$ (\emph{dotted}), and $44\degr$
(\emph{solid}).
Panel \textbf{b}: Ratio between positive and negative magnetic flux in
the polar region; the closer to one, the stronger the intermingling.
}
\label{fig:plot5}
\end{figure}

The surface flux transport simulations in Sect.\ \ref{simus} show that 
the intermediate range of originating latitudes (Case I, cf.\ Table
\ref{lats}) causes a higher degree of intermingling than the two cases
S and L.
To investigate this aspect in more detail, we analyse the relevant 
mechanisms in the simplified case of a single bipole evolving across
the surface under the effect of meridional flow, differential rotation
and supergranular diffusion.
For a meridional flow velocity of $v_M= 100\un{m\cdot s^{-1}}$, the
butterfly diagrams in Case S, I, and L have the \emph{mean} latitude of
bipole emergence $\bar{\lambda}_e= 21\degr$, $39\degr$, and $45\degr$,
respectively.
In each of the three cases, a single bipole is inserted onto a
(magnetically empty) stellar surface at the mean latitude of emergence. 
The initial tilt angle is $\bar{\lambda}_e/2$, whereas the initial 
total magnetic flux is in all cases the same ($1.5\cdot10^{23}\un{Mx}$).
The temporal evolution of the polar flux density and ratio of opposite 
polarities is shown Fig.\ \ref{fig:plot5}. 
The higher the (mean) latitude of emergence, the sooner the bipole 
reaches the polar region, and the stronger is the remaining polar flux,
due to the shorter time span available for flux annihilation as it is
transported to the poles.
The trailing flux region (here, of positive polarity) enters the polar 
region before the leading region (of negative polarity); when the 
latitude of emergence is small the timing when both polarities enter 
the polar region becomes similar.
In contrast, the higher the latitude of emergence the less
intermingling between magnetic flux of opposite polarity is obtained.
As the entire flux is pushed into the polar region, the flux ratio
eventually approaches unity.
Using the area under each curve as a measure for the degree of
intermingling, the results for the two lower mean latitudes of
emergence are similar, whereas the degree of intermingling for 
$\bar{\lambda}_e= 45\degr$ is overall considerably lower.

Since each bipole obeys Joy's law, the trailing flux region is located 
at somewhat higher latitudes than the preceding flux region of the
opposite polarity.
The meridional circulation then pushes both magnetic flux regions in 
this form toward the pole.
However, during the poleward transport the differential rotation 
however rotates the bipole.
This causes more flux of each polarity to be located at a common 
latitude, which is a requirement for intermingling to occur once the 
bipole is pushed into the polar region. 
If a bipole emerges at high latitudes, then the differential rotation 
has not enough time to act before its entry into the polar region.
Hence a higher degree of intermingling is obtained if the initial 
latitude of emergence is low.
This is in contrast to the polar flux density, which is found to 
decrease with the initial latitudes of emergence.
Consequently, there is a particular range of latitudes of emergence and 
poleward meridional flows, which results in both high degrees of 
intermingling and strong polar flux densities.
Optimal values occur when the timescales of both meridional flow and 
differential rotation are similar.

\label{lastpage}

\end{document}